\RequirePackage{fix-cm}

\documentclass[twocolumn,epjc3]{svjour3}  

\smartqed 

\RequirePackage{graphicx}
\usepackage{subcaption}
\usepackage{graphicx}
\usepackage{amsmath}
\usepackage{cite} 
\usepackage{mathrsfs}

\journalname{Eur. Phys. J. C}
\begin{document}

\title{Construction of an analytic multi-component accretion environment and its application to Kerr black hole imaging}

\author{Shiyang Hu\thanksref{addr1,addr2}
        \and
        Dan Li\thanksref{addr1,addr2}
        \and
        Chen Deng\thanksref{addr3}
        \and
        Kejian He\thanksref{addr4}
        \and
        Guansheng He\thanksref{addr1,addr2}
}

\thankstext{e1}{e-mail: husy$\_$arcturus@163.com}

\institute{School of Mathematics and Physics, University of South China,
           Hengyang, 421001, People's Republic of China \label{addr1}
           \and
           Hunan Provincial Key Laboratory of Mathematical Modeling and Scientific Computing, University of South China,
           Hengyang, 421001, People's Republic of China \label{addr2}
           \and
           School of Astronomy and Space Science, Nanjing University, 
           Nanjing, 210023, People's Republic of China \label{addr3}
           \and
           Department of Mechanics, Chongqing Jiaotong University,
           Chongqing, 400000, People's Republic of China \label{addr4}
}

\date{Received: date / Accepted: date}

\maketitle

\begin{abstract}
The construction of accretion environments is fundamental to black hole imaging. From a purely geometric perspective, we construct a novel analytic accretion environment comprising a geometrically thick disk, ring-like bumps with a Gaussian profile, and localized compact emission regions modeled by Gaussian distributions. This environment offers high flexibility, enabling independent adjustments of disk thickness, vertical structure, and the positions and morphologies of localized spots, thereby allowing it to qualitatively mimic high-energy astrophysical phenomena. Applying this model to the Kerr spacetime, we investigate the resulting images via radiative transfer and ray-tracing simulations. The results confirm the effectiveness of the numerical implementation and reveal novel observational signatures of Kerr black holes under multi-component illumination, including multiple bright spots and arc-like structures. This work provides a convenient and fully analytic framework for modeling accretion in curved spacetimes, and offers a new perspective on inferring accretion mechanisms and transient high-energy processes from image features.
\end{abstract}
\section{Introduction}
Imaging the supermassive black holes in the center of the Virgo cluster and that of the Milky Way provides a direct test of general relativity (GR) and opens a new avenue for investigating high-energy astrophysical processes in extreme gravitational environments \cite{2019ApJ...875L...1E,2022ApJ...930L..12E}. These landmark achievements have been made possible by the Event Horizon Telescope (EHT), which employs very long baseline interferometry (VLBI) to capture, with exceptional angular resolution, the synchrotron radiation emitted by the accreting plasma around these black holes. In essence, interpreting black hole images requires establishing a robust connection between observed image features and the underlying high-energy phenomena occurring in curved spacetime. Therefore, it is of critical importance that theoretical modeling of accretion environments incorporates a broad range of potential high-energy events to enable a comprehensive interpretation of observational data.

In astrophysics, the accretion environments of supermassive black holes are predominantly structured as accretion disks. These disks manifest in various forms depending on the physical conditions of the system \cite{Yuan:2014gma}. The standard thin disk model describes optically thick, geometrically thin disks where thermal emission dominates \cite{1973A&A....24..337S,1974ApJ...191..499P,2005ApJS..157..335L}. Radiatively inefficient accretion flows (RIAFs), including advection-dominated accretion flows (ADAFs), represent the opposite extreme, featuring optically thin, geometrically thick configurations with low radiative efficiency \cite{1977ApJ...214..840I,1994ApJ...428L..13N}. In strongly magnetized systems, magnetically arrested disks (MADs) can form \cite{2003PASJ...55L..69N,2021ApJ...910L..13E,2023Sci...381..961Y}, where the accumulation of magnetic flux near the black hole significantly modifies the accretion dynamics. While general relativistic magnetohydrodynamic (GRMHD) simulations \cite{2019ApJS..243...26P} provide a powerful framework for capturing these diverse accretion scenarios, resolving specific high-energy phenomena---such as magnetic reconnection, episodic flares, tidal disruption events, and jet launching---demands carefully tailored simulation setups and remains a significant challenge in practice. Moreover, these simulations are computationally demanding and pose significant obstacles to parameter space exploration. This motivates the search for analytic or semi-analytic alternatives, which offer a more tractable route for systematic investigations, especially for black hole imaging studies, where rapid exploration of the parameter space is essential.

In current theoretical simulations of black hole images, most investigations treat the light source around the black hole as either a spherically symmetric accretion flow or a steady-state accretion disk. In the former case, many studies have focused on the influence of the inner boundary, emissivity profile, and radiation frequency on the black hole shadow and the bright ring, revealing a robust alignment between the shadow contour and the critical curve \cite{2000ApJ...528L..13F,2019ApJ...885L..33N,2020EPJC...80..872Z,2023EPJC...83..277W,Guo:2021bwr,2023IJMPD..3250088H}. The latter case constitutes a broad topic with multiple branches. Based on the analytic, phenomenological, optically thin, geometrically thin, equatorial accretion disk established in \cite{2019PhRvD.100b4018G}, many scholars have explored the face-on images of various black holes, wormholes, and exotic compact objects, identifying imprints of spacetime parameters on the images \cite{He:2021htq,Li:2021riw,Zeng:2021dlj,Zeng:2021mok,Li:2021ypw,He:2022yse,Zeng:2022pvb,Wu:2024juj,Hu:2022lek,Peng:2021osd,Peng:2020wun,Meng:2025ivb,Wang:2024lte,Meng:2023htc,Wang:2023vcv,Huang:2025gia,Chen:2025ifv,Zeng:2023fqy}. Inspired by Luminet's celebrated hand-drawn black hole image and the Novikov-Thorne optically thick, geometrically thin accretion disk model \cite{1979A&A....75..228L}, a large body of work has examined the combined influence of observation inclination and spacetime parameters on the resulting images \cite{2019PhRvD.100f4011T,Hu:2023bzy,2021EPJC...81..885G,2022IJMPD..3150041L}. Chael et al. developed an analytic, optically thin, geometrically thin accretion disk model and a dynamical framework fitted to the time-averaged images from GRMHD simulations \cite{2021ApJ...918....6C}, which have played a key role in numerous black hole imaging studies \cite{Hu:2023pyd,2026ForPh..7470062L,Hou:2022eev,Zhang:2023bzv,Li:2024ctu,Zeng:2025pch,Zeng:2025tji,Wang:2025btn,Wan:2026xzs}. Some works have constructed geometrically thick accretion disk models to simulate polarization images of different black holes, providing guidance for revealing the magnetic field environment in strong gravitational fields \cite{Zhou:2025moa,Yang:2025byw,Hou:2024qqo,Huang:2024bar,Zhang:2023cuw,Zeng:2026ntc,Aslam:2025hgl,Wang:2025qpv,Zeng:2025kyv,Wan:2025gbm}. More interestingly, several scholars have built tilted accretion disk models from either magnetohydrodynamic or purely geometric perspectives, uncovering corresponding image features such as shadow erosion and the rotation of bright spots \cite{Hu:2023pyd,Chatterjee:2020eqc,White:2020cos,Hu:2024imp,Hu:2026rqa}. 

Evidently, whether for spherically symmetric accretion flows or accretion disks, these modeling approaches are highly efficient. They can not only qualitatively reveal the image features of different black holes, but also impose constraints on spacetime parameters based on EHT shadow observations. At the same time, their limitations are also apparent. If the accretion disk is not quiescent but contains additional internal components, if accretion does not take the form of a disk but instead forms other ring-like structures, or if further high-energy events occur within the disk, then the resulting image is expected to exhibit additional features that cannot be reproduced by a single accretion disk model.

To this end, we construct a rich accretion environment from a purely geometric perspective. This environment comprises a geometrically thick accretion disk, ring-like bumps, and localized compact emission regions, and it offers a high degree of flexibility to qualitatively mimic various potential high-energy phenomena around black holes. The remainder of this paper is organized as follows. In Sec. 2, we present the emission and absorption properties of the accretion environment, along with the dynamical framework, and describe the model parameters. In Sec. 3, we apply this accretion model to the Kerr black hole, investigate its properties through the emission profile, the dynamical profile, and the resulting images, and examine the relation between the model ingredients and the image features. In the final section, we present our conclusions and a brief discussion. Throughout this paper, we adopt geometrized units.
\section{Analytic multi-component accretion environment}
\subsection{Emission and absorption}
In the local coordinates of the accretion matter, $x=(t,r,\theta,\varphi)$, the dimensionless emissivity $j_{\nu}(r,\theta,\varphi)$ of our accretion environment model is composed of the attenuated background emission from the thick accretion disk $j_{\textrm{d}}(r,\theta)$, the Gaussian bumps $j_{\textrm{b}}(r,\theta)$, and the localized bright spots $j_{\textrm{s}}(r,\theta,\varphi)$, and is given by
\begin{equation}\label{1}
j_{\nu}(r,\theta,\varphi)=j_{0}\left[j_{1}j_{\textrm{d}}(r,\theta)+j_{2}j_{\textrm{b}}(r,\theta)+j_{3}j_{\textrm{s}}(r,\theta,\varphi)\right].
\end{equation}
Here, $j_{0}$ determines the overall radiation intensity, and $j_{i}$ ($i=1,2,3$) are the weight coefficients of the individual components. 

The term $j_{\textrm{d}}(r,\theta)$ describes the emission distribution of the geometrically thick accretion disk. This is an axisymmetric distribution model, in which the emission decays radially and follows a Gaussian profile about the equatorial plane\footnote{Interestingly, some modeling studies also introduce a tilt of the accretion disk \cite{Hu:2023pyd,Hu:2024imp,Hu:2026rqa,Chatterjee:2020eqc,White:2020cos}. However, it is worth noting that for a sufficiently evolved accretion disk, even if a tilted component is initially present, it tends to align toward the equatorial plane through the Bardeen-Petterson effect \cite{Liska:2018ayk,Liska:2019vne,Bardeen:1975zz,Nealon:2015jya}. Therefore, in constructing our model, we adopt the classical assumption of an equatorial accretion disk.}. Therefore, we can combine these two distributions to obtain $j_{\textrm{d}}(r,\theta)$. For the radial part, we adopt an exponential decay behavior, written in logarithmic space as
\begin{equation}\label{disk:r}
\ln\left[j_{\textrm{r}}\left(r\right)\right] = p_{1}\ln\left(\frac{r_{\textrm{eff}}}{r_{\textrm{in}}}\right) + p_{2}\left[\ln\left(\frac{r_{\textrm{eff}}}{r_{\textrm{in}}}\right)\right]^{2},
\end{equation}
where $p_{1}$ and $p_{2}$, as in \cite{2021ApJ...918....6C}, are used to finely control the decay rate of the emission; $r_{\textrm{in}}$ is the inner boundary of the geometrically thick accretion disk.

On top of the overall radial decay behavior, we introduce a radiative compensation to phenomenologically model a region where the radiative efficiency is locally enhanced or sustained despite the overall radial decline. Such a feature may arise from various physical mechanisms, including local energy injection due to turbulence or magnetic reconnection, or a persistent ring-like density enhancement within the disk. This compensation mechanism is realized through $r_{\textrm{eff}}$ in Eq. \eqref{disk:r}, written as
\begin{equation}\label{3}
r_{\textrm{eff}}=r-j_{\textrm{p}}w_{\textrm{p}}\left[\tanh(\frac{r-r_{\textrm{p}}}{w_{\textrm{p}}})-\tanh(\frac{r_{\textrm{in}}-r_{\textrm{p}}}{w_{\textrm{p}}})\right],
\end{equation}
where $r_{\textrm{p}}$, $w_{\textrm{p}}$, and $j_{\textrm{p}}$ denote the central position, the width, and the strength coefficient of the compensation, respectively. It can be seen that, through the hyperbolic tangent function, the compensation introduces a negative shift to the radial coordinate within the compensation region, thereby counteracting the radial decay. In general, this compensation introduces a delay in the radial decay of the emission profile, and we may also refer to this component as a plateau. In particular, the compensation can be switched off by setting $j_{\textrm{p}}=0$.

Next, we construct the Gaussian distribution function about the equatorial plane, written in logarithmic space as
\begin{equation}\label{disk:theta}
\ln\left[j_{\theta}\left(r,\theta\right)\right] = -\frac{\left(\theta-\pi/2\right)^{2}}{2\left(\sigma_{d\theta} \textrm{e}^{\beta r}\right)^{2}},
\end{equation}
where $\sigma_{d\theta}$ characterizes the disk thickness (opening angle), and $\beta$ controls the radial growth of the disk thickness. In particular, the term $\textrm{e}^{\beta r}$ makes the disk thickness vary nonlinearly (a linear variation would correspond to a disk with a fixed opening angle), with the expansion rate determined by the value of $\beta$. This allows the disk to exhibit a rich variety of geometric structures, with profiles that can resemble conical, flared, or teardrop-like shapes. Subsequently, we obtain the emission function of the geometrically thick accretion disk
\begin{equation}\label{disk:all}
\ln\left[j_{\textrm{d}}(r,\theta)\right] = \ln\left[j_{\textrm{r}}\left(r\right)\right] + \ln\left[j_{\theta}\left(r,\theta\right)\right].
\end{equation}

The accretion disk may host outward-propagating shock waves triggered by various mechanisms, such as density waves driven by accretion rate fluctuations, local energy injection from turbulence powered by the magnetorotational instability (MRI), or interactions between the disk and episodic flares or jets launched from the inner region. Although the propagation of such shock waves is a dynamical and highly complex fluid process, we can qualitatively model them mathematically using a phenomenological approach: a Gaussian bump $j_{\textrm{b}}(r,\theta)$ is superimposed on $j_{\textrm{d}}(r,\theta)$, and is expressed in logarithmic space as
\begin{equation}\label{4}
\ln\left[j_{\textrm{b}}(r,\theta)\right]=-\frac{\left(r-r_{\textrm{b}}\right)^{2}}{2\sigma_{br}^{2}}-\frac{\left(\theta-\pi/2\right)^{2}}{2\sigma_{b\theta}^{2}}.
\end{equation}
Here, $r_{\textrm{b}}$ denotes the radial position of the bump, while $\sigma_{br}$ and $\sigma_{b\theta}$ characterize its radial and angular extents, respectively. It is worth noting that the overall structure of the shock wave is axisymmetric: its projection onto the equatorial plane is a ring with non-negligible thickness, while its projection onto the plane containing the black hole spin axis is predominantly elliptical or bow-shaped. These geometric features are naturally suited to being characterized by the parameters of a Gaussian distribution. Specifically, when $\sigma_{br}$ is large and $\sigma_{b\theta}$ is small, the cross-section of the shock resembles a pebble; conversely, it takes the form of a bow shock.

Localized compact emission regions may also appear within the accretion disk, arising from collisions between material streams, tidal disruption events (TDE), companion stars, or magnetic reconnection episodes. Such features can be qualitatively modeled by adding a point source of non-negligible size. Moreover, due to the gravitational field of the central compact object, such a point source is often subject to differential tidal forces and becomes distorted into a ``pear-like'' shape, with its head oriented toward the central object. This deformation behavior of the point source is well suited to being characterized by a Gaussian distribution. Therefore, in logarithmic space, the emissivity of the localized spot incorporating these effects can be written as\footnote{In numerical implementations, the azimuthal term must be written as $1-\cos(\varphi-\varphi_{\textrm{s}})$ instead of $\left(\varphi-\varphi_{\textrm{s}}\right)^{2}$ to correctly handle the $2\pi$ periodicity of the coordinate $\varphi$. Using the quadratic form would produce artificial deviations when the spot lies near the $\varphi=0$ boundary, leading to incorrect spot localization in the code.}
\begin{equation}\label{5}
\ln\left[j_{\textrm{s}}(r,\theta,\varphi)\right]=-\frac{\left(r-r_{\textrm{s}}\right)^{2}}{2\sigma_{sr}^{2}}-\frac{\left(\theta-\theta_{\textrm{s}}\right)^{2}}{2\sigma_{s\theta}^{2}}-\frac{\left(\varphi-\varphi_{\textrm{s}}\right)^{2}}{2\sigma_{s\varphi}^{2}},
\end{equation}
where $(r_{\textrm{s}},\theta_{\textrm{s}},\varphi_{\textrm{s}})$ denotes the position of the spot in the black hole coordinates, and $\sigma_{sr}$, $\sigma_{s\theta}$, and $\sigma_{s\varphi}$ collectively control its size and asymmetric deformation. Notably, our model offers excellent extensibility. For instance, if multiple spots are to be introduced, one simply writes the corresponding $j_{\textrm{s}}(r,\theta,\varphi)$ for each additional spot following the form of \eqref{5} and then incorporate them into $j_{\nu}(r,\theta,\varphi)$. In addition, it is worth noting that although both the shock wave and the localized spot adopt a Gaussian form, their contributions to the accretion environment are distinct: the former is a large-scale structure appearing on the accretion disk and present across the entire azimuthal range, whereas the latter is a localized contribution existing only in the vicinity of $\varphi_{\textrm{s}}$.

The absorption function $\alpha_{\nu}(r,\theta,\varphi)$ corresponding to the emission model \eqref{1} is given by
\begin{eqnarray}\label{9}
\alpha_{\nu}(r,\theta,\varphi)=\alpha_{0}\left[\alpha_{1}\alpha_{\textrm{d}}(r,\theta)+\alpha_{2}\alpha_{\textrm{b}}(r,\theta)+\alpha_{3}\alpha_{\textrm{s}}(r,\theta,\varphi)\right].
\end{eqnarray}
Here, $\alpha_{\textrm{d}}(r,\theta)$, $\alpha_{\textrm{b}}(r,\theta)$, and $\alpha_{\textrm{s}}(r,\theta,\varphi)$ denote the absorption of the accretion disk, the bump, and the localized spot, respectively. For convenience, these components follow the same analytic forms as their emission counterparts, although the model parameters may take different values. Moreover, by adjusting $\alpha_{i}$ ($i=0,1,2,3$), one can set the absorption strength of each component and thereby control the optical thickness of the medium. For example, setting $\alpha_{0}=0$ corresponds to the optically thin limit.
\subsection{Dynamics}
The motion of the accreting matter introduces Doppler effects on the specific intensity of light and is therefore an essential component of the accretion model. We consider a general axisymmetric spacetime whose line element is given by
\begin{eqnarray}\label{10}
\textrm{d}s^{2}=g_{tt}\textrm{d}t^{2}+g_{rr}\textrm{d}r^{2}+g_{\theta\theta}\textrm{d}\theta^{2}+g_{\varphi\varphi}\textrm{d}\varphi^{2}+2g_{t\varphi}\textrm{d}t\textrm{d}\varphi.
\end{eqnarray}
In this spacetime, we introduce the Zero Angular Momentum Observer (ZAMO), whose tetrad basis is given by \cite{Hou:2022eev,Huang:2024bar,Cunha:2016bpi,Wang:2021ara,Cao:2023ppv,M:2022pex,Zhong:2021mty}
\begin{eqnarray}
\hat{e}_{0} &=& \frac{1}{\kappa}\left(\partial_{t}+\Omega\partial_{\varphi}\right), \label{11} \\
\hat{e}_{1} &=& \frac{1}{\sqrt{g_{rr}}}\partial_{r}, \label{12} \\
\hat{e}_{2} &=& \frac{1}{\sqrt{g_{\theta\theta}}}\partial_{\theta}, \label{13} \\
\hat{e}_{3} &=& \frac{1}{\sqrt{g_{\varphi\varphi}}}\partial_{\varphi}, \label{14}
\end{eqnarray}
where $\kappa$ and $\Omega$ are the lapse function and the angular velocity of the frame dragging, respectively, and are expressed as
\begin{equation}\label{15}
\kappa = \sqrt{-g_{tt}+\frac{g_{t\varphi}^{2}}{g_{\varphi\varphi}}},
\end{equation}
and
\begin{equation}\label{16}
\Omega = -\frac{g_{t\varphi}}{g_{\varphi\varphi}}.
\end{equation}
The four-velocity of the accreting material described in the local frame of the axisymmetric spacetime is then written as
\begin{equation}\label{17}
u^{\mu}=\Gamma\left(\hat{e}_{0}^{\mu}+\hat{v}^{i}\hat{e}_{i}^{\mu}\right).
\end{equation}
Here, $\hat{v}^{i}=(\hat{v}^{r},\hat{v}^{\theta},\hat{v}^{\varphi})$ is the three-velocity of the accreting material measured by the ZAMO, and $\Gamma$ is the Lorentz factor,
\begin{equation}\label{18}
\Gamma = \frac{1}{\sqrt{1-\left(\hat{v}^{r}\right)^{2}-\left(\hat{v}^{\theta}\right)^{2}-\left(\hat{v}^{\varphi}\right)^{2}}}.
\end{equation}
From the tetrad basis \eqref{11}--\eqref{14}, the four-velocity components of the accreting material can be expressed as
\begin{eqnarray}
u^{t} &=& \frac{\Gamma}{\kappa}, \label{19} \\
u^{r} &=& \frac{\Gamma \hat{v}^{r}}{\sqrt{g_{rr}}}, \label{20} \\
u^{\theta} &=& \frac{\Gamma \hat{v}^{\theta}}{\sqrt{g_{\theta\theta}}}, \label{21} \\
u^{\varphi} &=& \Gamma\left(\frac{\Omega}{\kappa}+\frac{\hat{v}^{\varphi}}{\sqrt{g_{\varphi\varphi}}}\right). \label{22}
\end{eqnarray}

We assume that the accreting material is injected from large radii toward the central object along conical surfaces, i.e., $\hat{v}^{\theta}=0$, with its radial velocity increasing monotonically as it approaches the center. The radial velocity $\hat{v}^{r}$ and the azimuthal velocity $\hat{v}^{\varphi}$ of the material in the ZAMO frame are then prescribed. For the radial component, we adopt a power-law distribution
\begin{equation}\label{23}
\hat{v}^{r}(r) = -V_{\textrm{max}}\left(\frac{r_{\textrm{in}}}{r}\right)^{p_{3}}.
\end{equation}
Here, $V_{\textrm{max}}$ is the maximum radial velocity of the accretion flow, i.e., the maximum radial velocity allowed for the accreting material; $p_{3}$ controls the acceleration rate of the material; the negative sign indicates that the velocity is directed inward. It can be seen that, in the radial direction, the accreting material undergoes global acceleration, indicating that the material continuously converts gravitational potential energy into its own kinetic energy.

For the azimuthal component, we assume that at large radii the accretion flow is in nearly circular orbital motion with a relatively low angular velocity. As the material moves inward, the azimuthal velocity gradually increases. However, near the inner boundary of the accretion model, the azimuthal motion is artificially suppressed so that the flow becomes radially dominated. This setup is consistent with the astrophysical context: during the accretion process, viscous torques continuously transport angular momentum outward, so that when the material reaches the inner region of the accretion disk, the centrifugal force arising from its excessively low angular momentum is insufficient to balance gravity, causing the azimuthal motion to be suppressed and the accretion to become radially dominated. The analytic expression designed to capture this physical picture is given by
\begin{equation}\label{24}
\hat{v}^{\varphi}(r) = \frac{\psi\sqrt{1/r}}{1+\lambda\left(1/r\right)},
\end{equation}
where $\psi$ sets the overall rotational speed of the material, and $\lambda$ regulates the suppression of rotation in the inner region. A larger value of $\lambda$ leads to a more efficient suppression of the orbital motion, so that the radial infall becomes increasingly dominant. It can be seen that when $r$ is large, the suppression induced by $\lambda$ is almost negligible, and the azimuthal velocity scales as $\propto \sqrt{1/r}$, consistent with the Keplerian velocity scaling; when $r$ is small, the suppression effect of $\lambda$ becomes significant.

Substituting $\hat{v}^{r}$ and $\hat{v}^{\varphi}$ into Eqs. \eqref{19}--\eqref{22} yields the four-velocity $u^{\mu}$ of the accreting material in the local frame of the axisymmetric spacetime. Importantly, the resulting four-velocity naturally satisfies the normalization condition and consistently incorporates the frame-dragging effect. It is worth noting that setting $\hat{v}^{\theta}=0$ is a classic and phenomenological operation for treating the disk dynamics within the ballistic approximation \cite{Hou:2023bep}. It ensures mathematical simplicity while highlighting the fact that the poloidal velocity of the accreting material is typically much smaller than its radial velocity, and it can efficiently capture the picture of material accelerating inward toward the central object along conical surfaces. In fact, our model is highly flexible, as it supports specifying arbitrary physical three-velocity components $\left(\hat{v}^{r},\hat{v}^{\theta},\hat{v}^{\varphi}\right)$ and then obtaining the corresponding four-velocity through Eqs. \eqref{19}--\eqref{22}, thereby allowing us to describe various possible kinematic behaviors of the accreting material.
\section{Applications in Kerr spacetime}
\subsection{Metric}
In Boyer-Lindquist coordinates with the spacelike signature $(-,+,+,+)$, the dimensionless line element of the Kerr spacetime is written as 
\begin{eqnarray}\label{25}
\textrm{d}s^{2} &=& g_{\mu\nu}\dot{x}^{\mu}\dot{x}^{\nu} \nonumber \\
&=&-\left(1-\frac{2r}{\Sigma}\right)\textrm{d}t^{2}-\frac{4ar\sin^{2}\theta}{\Sigma}\textrm{d}t\textrm{d}\varphi+\frac{\Sigma}{\Delta}\textrm{d}r^{2} +\Sigma \textrm{d}\theta^{2} \nonumber \\
&& +\sin^{2}\theta\left(r^{2}+a^{2}+\frac{2a^{2}r\sin^{2}\theta}{\Sigma}\right)\textrm{d}\varphi^{2},
\end{eqnarray}
where $g_{\mu\nu}$ is the covariant metric tensor, and $a$ is the dimensionless spin parameter. The quantities $\Sigma$ and $\Delta$ are defined as
\begin{eqnarray}\label{26}
\Sigma = r^{2}+a^{2}\cos^{2}\theta,
\end{eqnarray}
\begin{equation}\label{27}
\Delta = r^{2}-2r+a^{2}.
\end{equation}
From $\Delta = 0$, the event horizon radius is obtained as $r_{\textrm{eh}}=1+\sqrt{1-a^{2}}$. In the following calculations, we set the inner boundary of the accretion model to coincide with the event horizon, i.e., $r_{\textrm{in}}=r_{\textrm{eh}}$. This choice is consistent with the astrophysical context, as for low-luminosity supermassive black holes, the millimeter-wavelength electromagnetic radiation near the event horizon is optically thin \cite{2021ApJ...920..155B}.
\subsection{Accretion model profiles}
\subsubsection{Emission profiles in one-dimensional plane}
\begin{figure*}
\centering                   
\includegraphics[width=15cm]{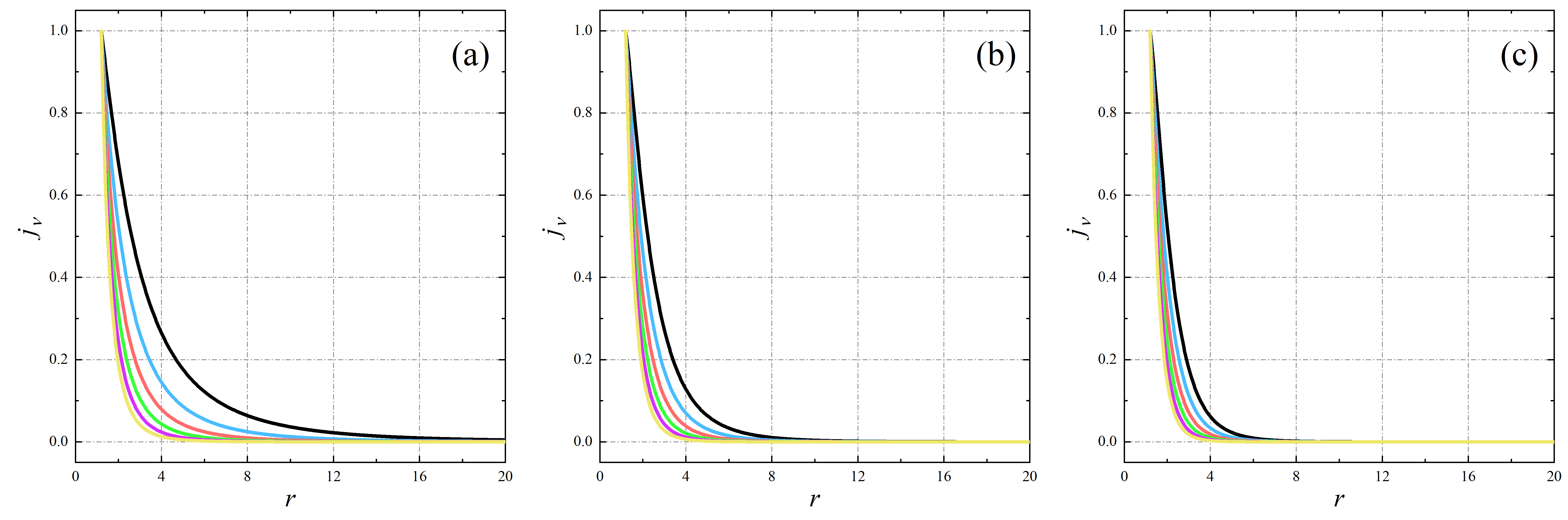}
\caption{Emission of the accretion model as a function of $r$ for different values of $p_{1}$ and $p_{2}$. Here, the inner boundary of the accretion model, $r_{\textrm{in}}$, coincides with the event horizon corresponding to $a=0.94$. Only the contribution of $j_{\textrm{d}}(r,\theta)$ is considered, and the plateau is turned off ($r_{\textrm{eff}}=r$). In each panel, from black to yellow, $p_{1}$ varies from $-0.5$ to $-3$ in steps of $-0.5$. From left to right, $p_{2}$ takes the values $-0.5$, $-1$, and $-1.5$, respectively.}      
\label{fig1}                 
\end{figure*}

We first display the radial emission profile of different combinations of accretion components in the equatorial plane. Fig. \ref{fig1} shows the influence of $p_{1}$ and $p_{2}$ on the disk emission when only $j_{\textrm{d}}(r,\theta)$ is present and the compensation is absent ($r_{\textrm{eff}}=r$). The emission peaks at the inner boundary of the accretion disk and decays monotonically with increasing $r$. The decay rate depends on $p_{1}$ and $p_{2}$: increasing the absolute value of either parameter significantly enhances the radial decay. 

\begin{figure*}
\centering                   
\includegraphics[width=15cm]{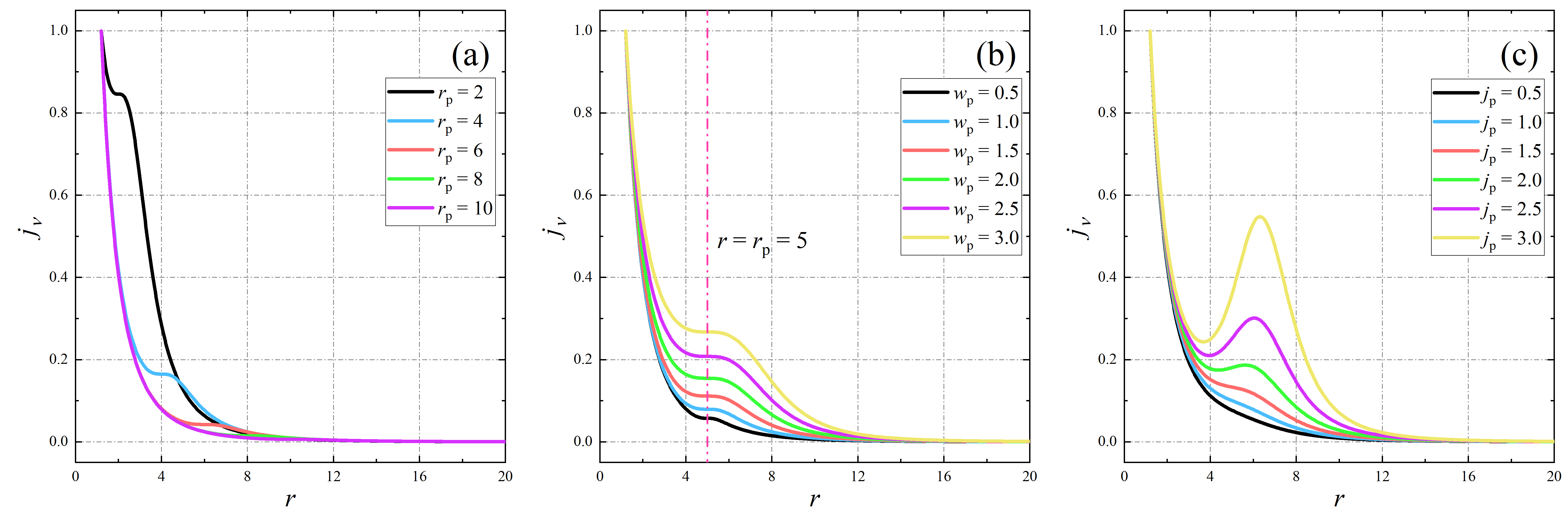}
\caption{Emission $j_{\nu}$ as a function of $r$ for different compensation parameters. Here, We fix $p_{1}=-1.5$ and $p_{2}=-0.5$, with $w_{\textrm{p}}=j_{\textrm{p}}=1$ for panel (a), $r_{\textrm{p}}=5$ and $j_{\textrm{p}}=1$ for panel (b), and $r_{\textrm{p}}=5$ and $w_{\textrm{p}}=2$ for panel (c).}      
\label{fig2}                 
\end{figure*}

We fix $p_{1}=-1.5$ and $p_{2}=-0.5$ and examine the influence of the compensation parameters $r_{\textrm{p}}$, $j_{\textrm{p}}$, and $w_{\textrm{p}}$ on the emission, as shown in Fig. \ref{fig2}. From panel (a), we observe that $r_{\textrm{p}}$ not only determines the position of the plateau, but also affects its visibility: a larger $r_{\textrm{p}}$ tends to obscure the compensation. This is because the emission decays rapidly with radius; at larger $r$, the curve becomes nearly flat, making the plateau difficult to identify. Increasing the width parameter $w_{\textrm{p}}$ not only broadens the compensation but also raises its emission level, as shown in panel (b). The effect brought by $j_{\textrm{p}}$ is primarily to enhance the contrast of the compensation. As $j_{\textrm{p}}$ increases, the compensation transitions from being nearly negligible (black curve) to a prominent bump-like feature (yellow curve), as shown in panel (c). In fact, according to our formulation, when $j_{\textrm{p}}=1$, the compensation is introduced as a horizontal extension; when $j_{\textrm{p}} > 1$, it takes on a raised shape; and when $j_{\textrm{p}} < 1$, the compensation exhibits a downward slope.

\begin{figure*}
\centering                   
\includegraphics[width=10cm]{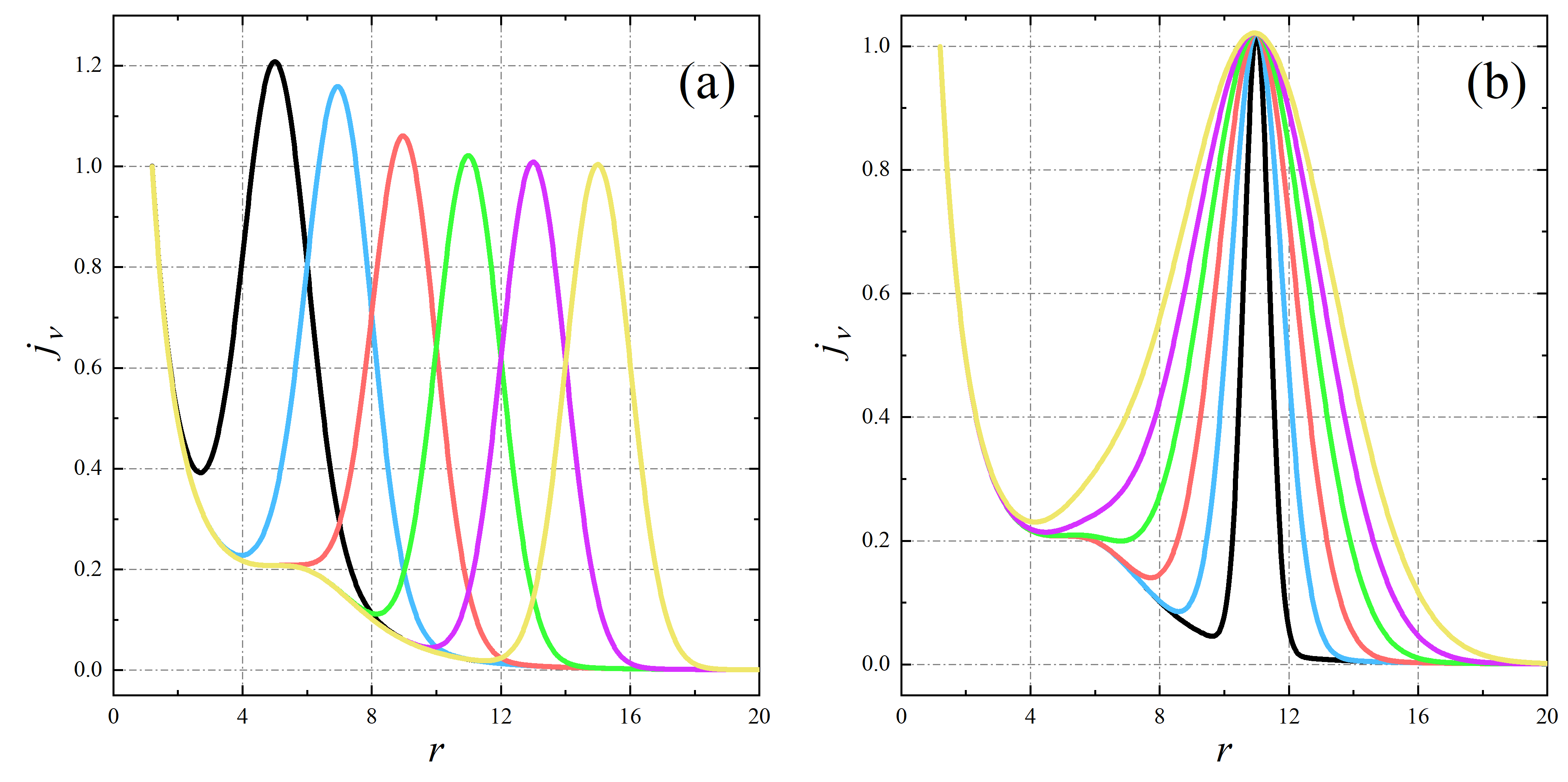}
\caption{Influence of Gaussian bumps $j_{\textrm{b}}(r,\theta)$ with different morphologies on $j_{\nu}(r,\theta,\varphi)$. In panel (a), from the black to the yellow curve, $r_{\textrm{b}}$ increases from $5$ to $15$ in steps of $2$, with $\sigma_{br}=1$ fixed. In panel (b), the bump position is fixed at $r_{\textrm{b}}=11$, while $\sigma_{br}$ increases from $0.4$ to $2.4$ in steps of $0.4$, from the black to the yellow curve.}      
\label{fig3}                 
\end{figure*}
Next, keeping $p_{1}=-1.5$, $p_{2}=-0.5$, $r_{\textrm{p}}=5$, $j_{\textrm{p}}=1$, and $w_{\textrm{p}}=2.5$ unchanged, we superimpose $j_{\textrm{b}}(r,\theta)$ onto the disk emission, i.e., $j_{\nu}(r,\theta,\varphi)=j_{\textrm{d}}(r,\theta)+j_{\textrm{b}}(r,\theta)$, to illustrate the effects of Gaussian bumps with different positions and widths in the accretion model. As shown in Fig. \ref{fig3}, $r_{\textrm{b}}$ effectively shifts the central position of the bump, while $\sigma_{br}$ controls its width. When the bump is located at the plateau position, the plateau feature is masked and replaced by a superposed emission bump, as illustrated by the black curve in panel (a). Similarly, when the bump has a relatively large $\sigma_{br}$ and is placed close to the plateau, the plateau can also be obscured, as shown by the yellow and purple curves in panel (b). In particular, when the bump is superposed on the plateau, the local emission can exceed the emission at the event horizon.

\begin{figure*}
\centering                   
\includegraphics[width=5cm]{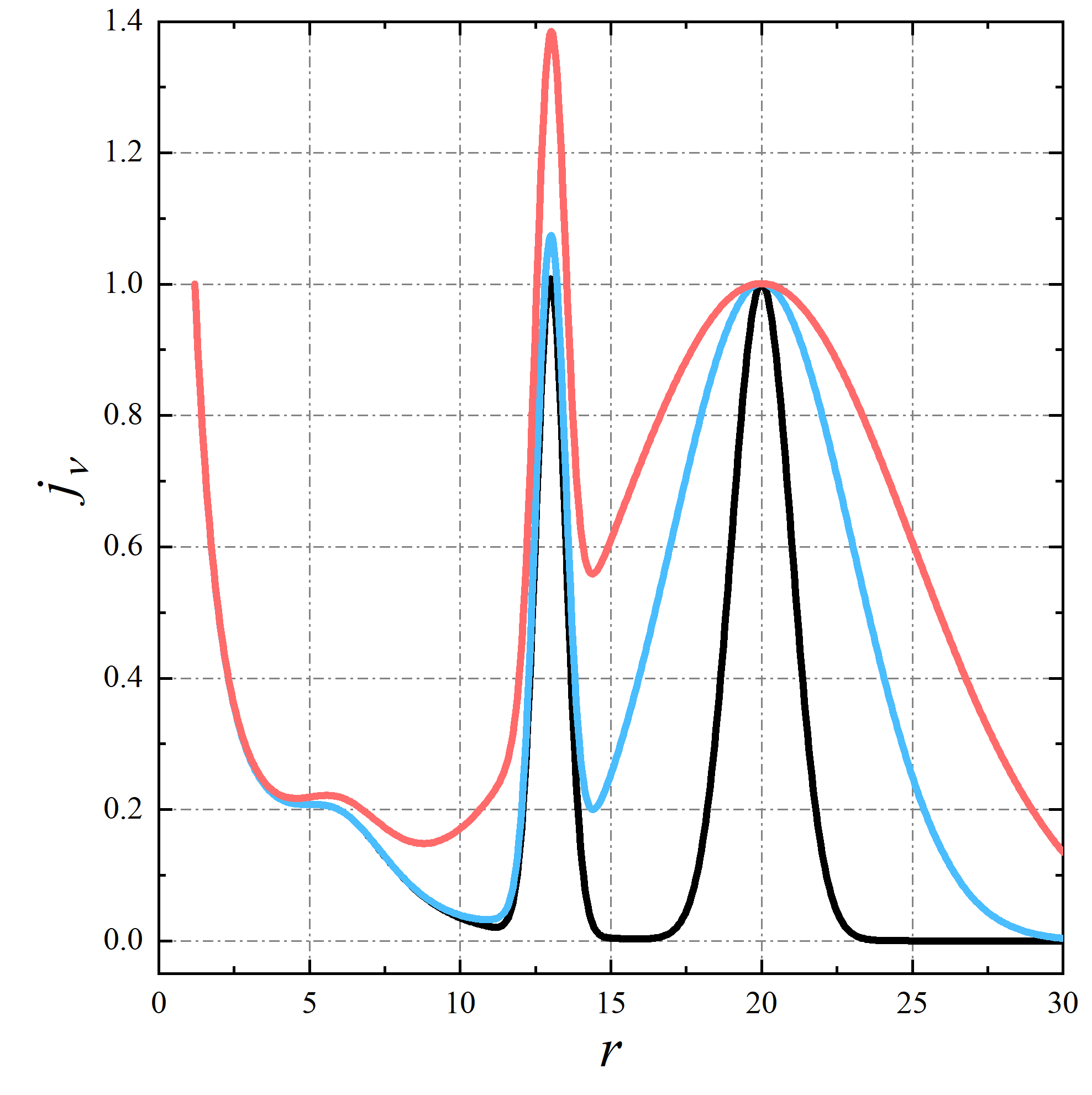}
\caption{Emission $j_{\nu}(r,\theta,\varphi)$ of the accretion model as a function of $r$ for different parameters of the localized spot. Here, the azimuthal angle examined is $\pi/4$. The black, blue, and red curves correspond to $\sigma_{sr}=1$, $3$, and $5$, respectively.}      
\label{fig4}                 
\end{figure*}
Superimposed on the base model $j_{\nu}(r,\theta,\varphi)=j_{\textrm{d}}(r,\theta)+j_{\textrm{b}}(r,\theta)$, we add a localized spot $j_{\textrm{s}}(r,\theta,\varphi)$ located at $(r_{\textrm{s}},\theta_{\textrm{s}},\varphi_{\textrm{s}})=(20,\pi/2,\pi/4)$. For $j_{\textrm{d}}(r,\theta)$, we adopt the same background as in Fig. \ref{fig3}, and fix $r_{\textrm{b}}=13$ and $\sigma_{br}=0.5$. The corresponding results are presented in Fig. \ref{fig4}. We note that introducing a localized spot is equivalent to adding a bump in the radial direction of the emission, whose width is determined by $\sigma_{sr}$. In particular, the localized spot can superpose with other components within the model, giving rise to enhanced emission. The sharp peak shown by the red curve results from the combined contribution of the Gaussian bump and the broad radial extent of the localized spot.
\subsubsection{Emission profiles in two-dimensional plane}
\begin{figure*}
\centering                   
\includegraphics[width=15cm]{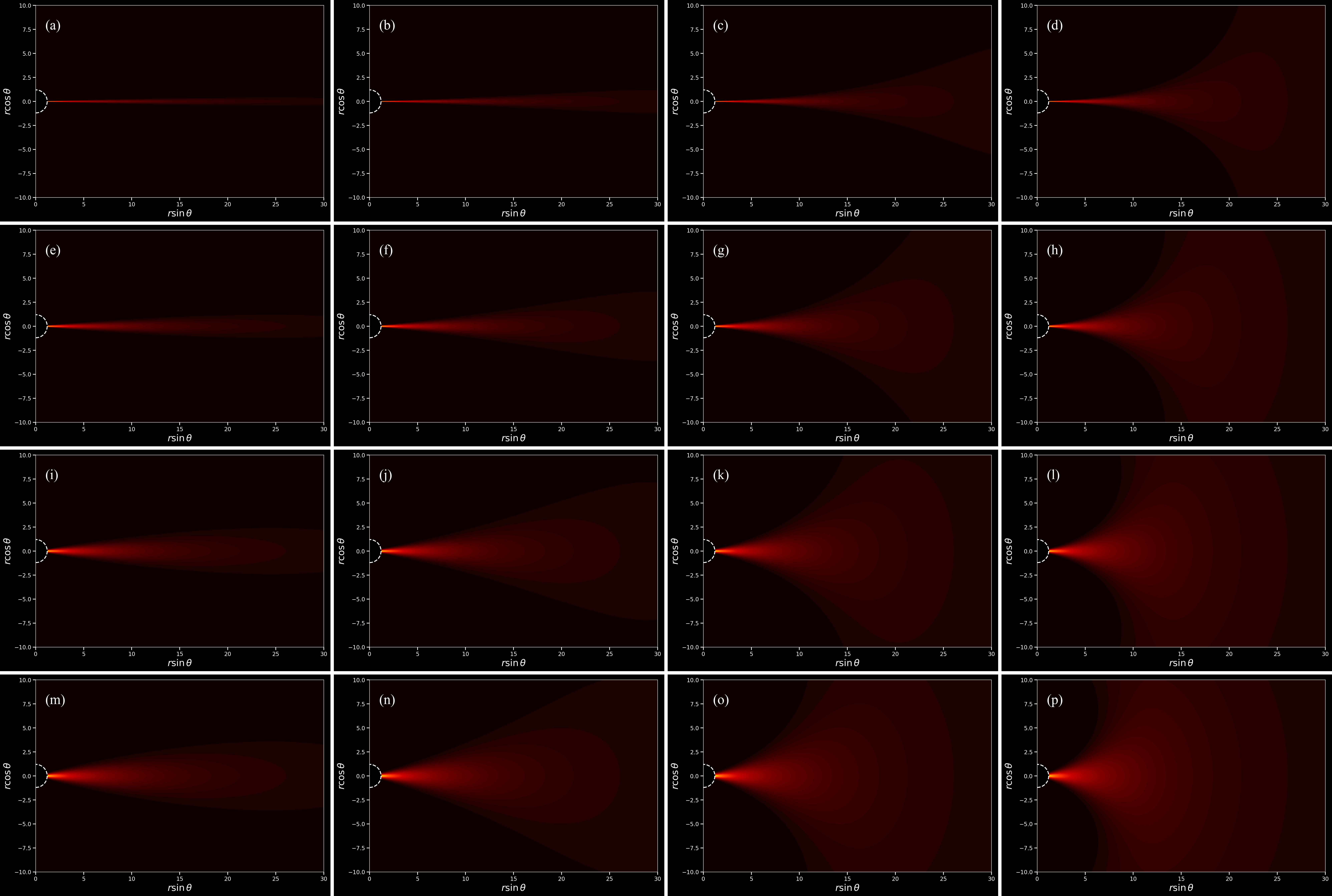}
\caption{Distributions of $j_{\nu}(r,\theta,\varphi)=j_{\textrm{d}}(r,\theta)$ in the two--dimensional plane $r\sin\theta$ versus $r\cos\theta$ for different values of the disk thickness parameters. From left to right, $\beta=0.01$, $0.05$, $0.1$, and $0.15$; from top to bottom, $\sigma_{d\theta}=0.01$, $0.03$, $0.06$, and $0.09$. All images are plotted in a gamma color scale with index $0.6$, and the plateau contribution is turned off ($j_{\textrm{p}}=0$).}      
\label{fig5}                 
\end{figure*}
From Figs. \ref{fig1} through \ref{fig4}, it is evident that our accretion model offers substantial flexibility to accommodate a wide range of complex accretion environments. However, the story does not end here. We now extend our inspection to the two--dimensional plane to further reveal the broad compatibility of the model. Fixing $p_{1}=-0.5$ and $p_{2}=-0.25$, we examine the influence of $\sigma_{d\theta}$ and $\beta$ on the disk thickness for the case $j_{\nu}(r,\theta,\varphi)=j_{\textrm{d}}(r,\theta)$ with no plateau contribution, as shown in Fig. \ref{fig5}. We find that when both $\beta$ and $\sigma_{d\theta}$ are very small, the disk exhibits almost no geometric thickness, as seen in panel (a). As $\beta$ increases, the disk becomes noticeably thicker, with its upper and lower surfaces extending toward the polar directions, resulting in a flared morphology. When the baseline thickness parameter $\sigma_{d\theta}$ is increased, the disk becomes thicker overall, and the flaring effect induced by $\beta$ becomes more prominent, manifesting primarily as an inward shift of the radius at which significant thickening begins.

\begin{figure*}
\centering                   
\includegraphics[width=15cm]{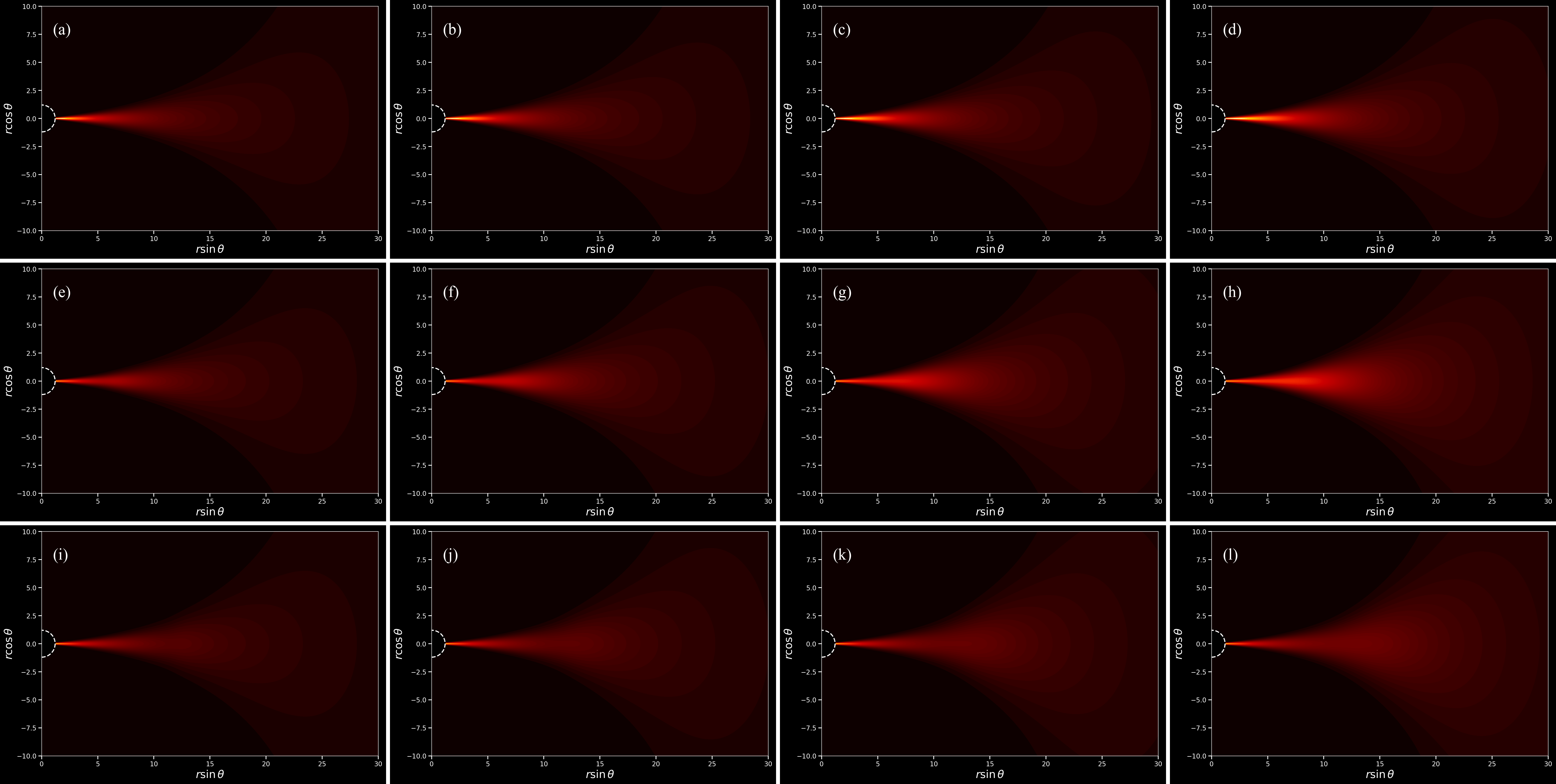}
\caption{Distributions of $j_{\nu}(r,\theta,\varphi)=j_{\textrm{d}}(r,\theta)$ in the two--dimensional plane $r\sin\theta$ versus $r\cos\theta$ for different plateau parameters. From left to right, the plateau width $w_{\textrm{p}}$ increases from $1$ to $4$ in steps of $1$; from top to bottom, the plateau position $r_{\textrm{p}}$ is set to $1.5$, $6$, and $12$, respectively. All other parameters, as well as the gamma color scale, are the same as in panel (g) of Fig. \ref{fig5}.}      
\label{fig6}                 
\end{figure*}
Building upon panel (g) of Fig. \ref{fig5}, we introduce the plateau with $j_{\textrm{p}}=1$, as shown in Fig. \ref{fig6}, where from top to bottom the plateau moves outward, and from left to right the plateau becomes wider. It is evident that the plateau effectively provides emission compensation to the disk. For instance, in the first row, as $w_{\textrm{p}}$ increases, the emission in the inner region of the accretion disk becomes progressively stronger. However, this compensation gradually diminishes as the plateau moves to larger radii, as can be seen in the bottom row.

\begin{figure*}
\centering                   
\includegraphics[width=15cm]{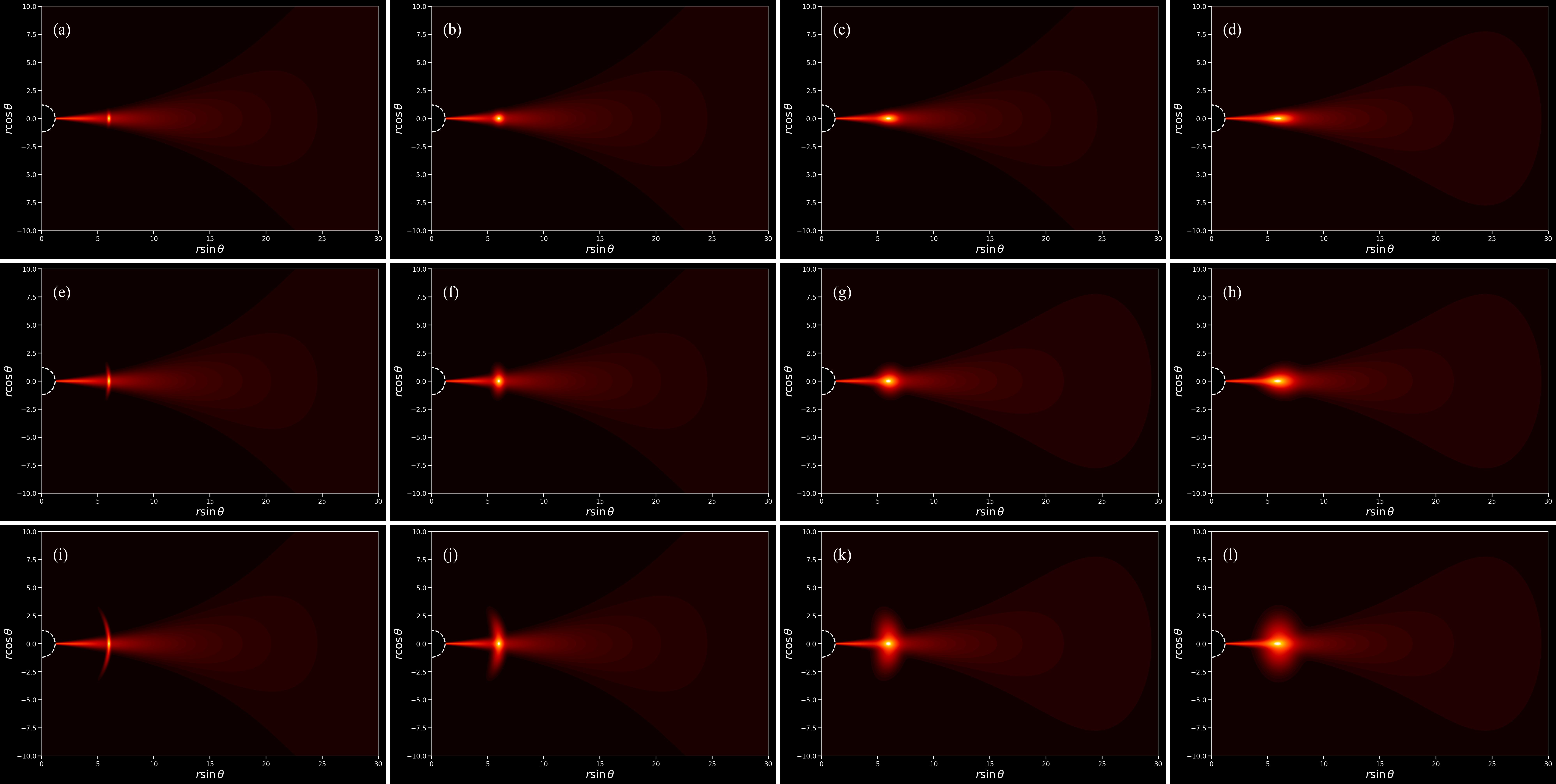}
\caption{Distributions of $j_{\nu}(r,\theta,\varphi)$ after introducing Gaussian bumps of different morphologies into the accretion disk. From left to right, $\sigma_{br}=0.1$, $0.3$, $0.6$, and $0.9$; from top to bottom, $\sigma_{b\theta}=0.05$, $0.1$, and $0.2$, respectively. The bump center is fixed at $r_{\textrm{b}}=6$, and all other parameters are the same as in panel (c) of Fig. \ref{fig6}.}      
\label{fig7}                 
\end{figure*}
Building upon panel (c) of Fig. \ref{fig6}, we examine the influence of the Gaussian bump on the emission, as shown in Fig. \ref{fig7}. The results show that when $\sigma_{br}$ is small, as in the first column, the Gaussian bump resembles an outward-propagating shock-like structure, which extends toward the polar directions as $\sigma_{b\theta}$ increases, and can even exceed the upper and lower surfaces of the disk. Increasing $\sigma_{br}$ broadens the bump in the radial direction, causing $j_{\textrm{b}}(r,\theta)$ to transition from a crescent-like shape to an elliptical one. It is worth noting that a plateau with $j_{\textrm{p}} > 1$ can also mimic the bump effect, although it offers less flexibility than $j_{\textrm{b}}(r,\theta)$.

\begin{figure*}
\centering                   
\includegraphics[width=15cm]{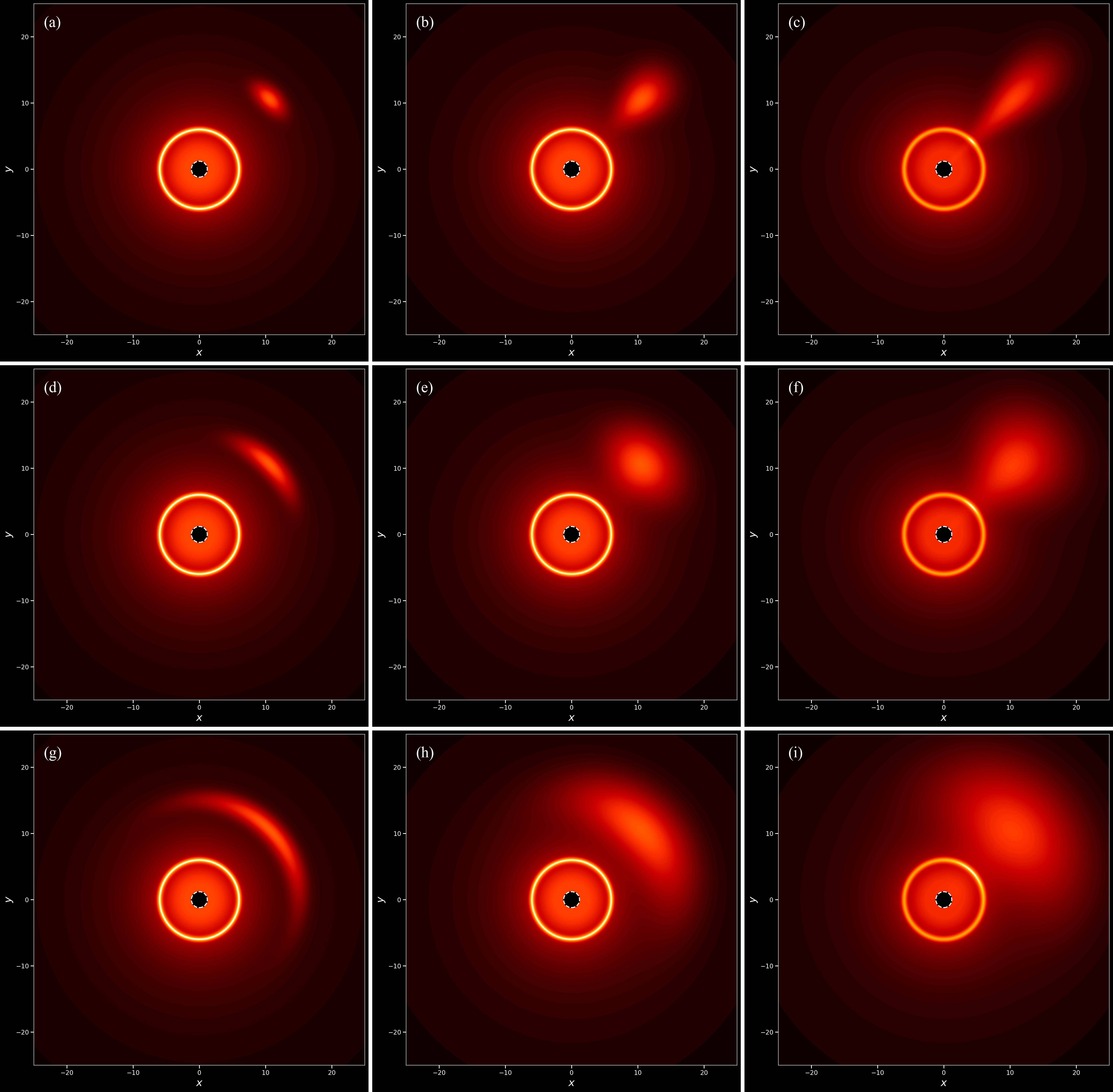}
\caption{Distributions of $j_{\nu}(r,\theta,\varphi)$ in the equatorial plane of the black hole for different localized spot parameters. From left to right, $\sigma_{sr}=1$, $3$, and $5$, respectively; from top to bottom, $\sigma_{s\varphi}=\pi/36$, $\pi/18$, and $\pi/9$, respectively. All other parameters are the same as in panel (f) of Fig. \ref{fig7}.}      
\label{fig8}                 
\end{figure*}
Finally, building upon panel (f) of Fig. \ref{fig7}, we examine the influence of the localized spot on the disk emission through Fig. \ref{fig8}. The spot position is fixed at $(r_{\textrm{s}},\theta_{\textrm{s}},\varphi_{\textrm{s}})= (15,\pi/2,\pi/4)$. We find that when $\sigma_{s\varphi}$ is small, increasing $\sigma_{sr}$ stretches the spot in the radial direction, deforming it into a teardrop shape with its tip pointing toward the black hole, which closely resembles the scenario of a companion star being accreted by the black hole. Conversely, when $\sigma_{sr}$ is small, increasing $\sigma_{s\varphi}$ deforms the spot into a crescent shape; further increasing $\sigma_{sr}$ makes the spot appear as a large-scale emission blob, as shown in panel (i). It is worth noting that the localized spot differs from the Gaussian bump, since the former is localized in azimuth while the latter contributes over the entire azimuthal direction.

\begin{figure*}
\centering                   
\includegraphics[width=15cm]{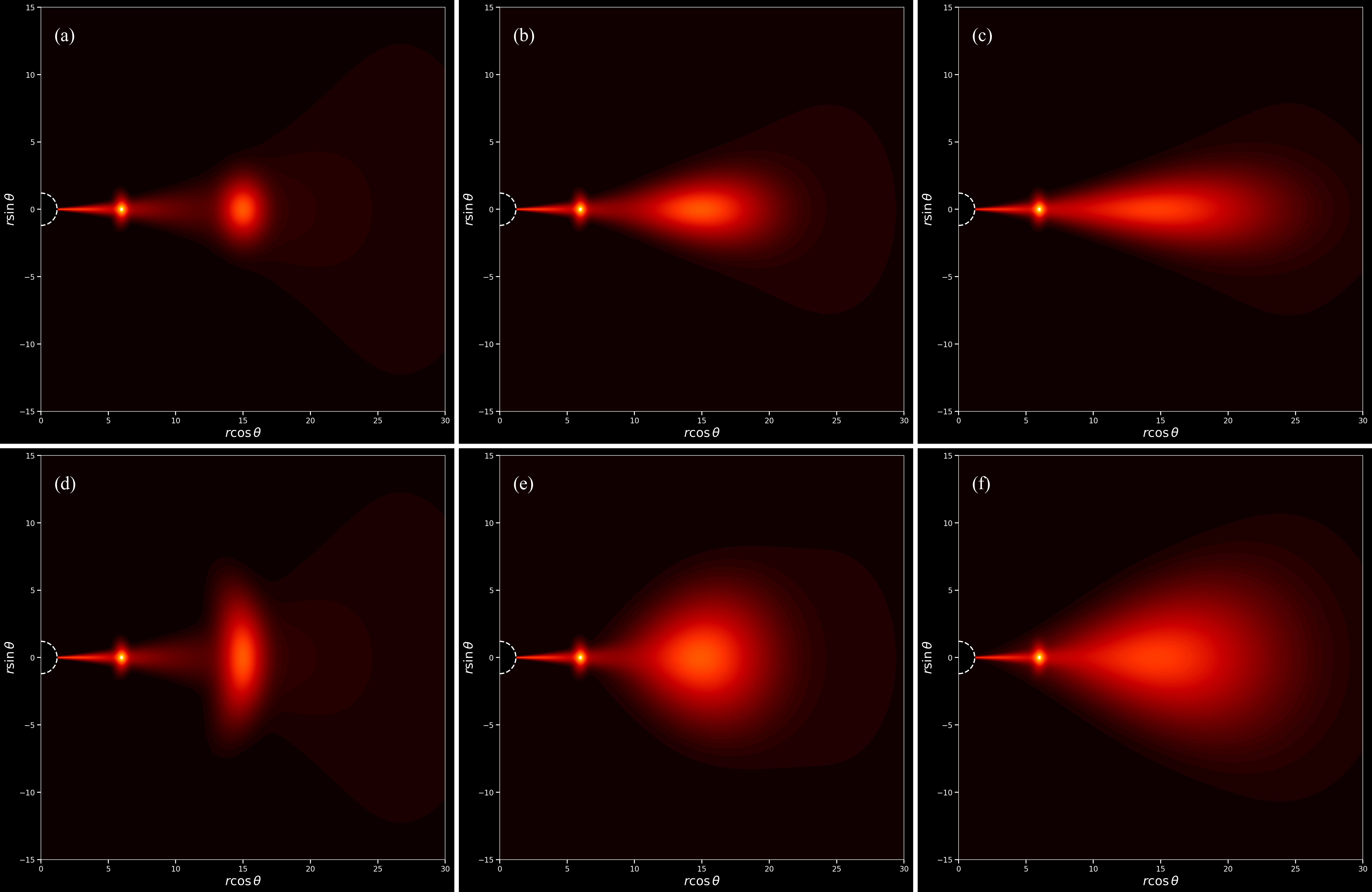}
\caption{Distributions of $j_{\nu}(r,\theta,\varphi)$ in the plane containing the black hole spin axis for different localized spot parameters. From left to right, $\sigma_{sr}=1$, $3$, and $5$, respectively; the first and second rows correspond to $\sigma_{s\theta}=\pi/36$ and $\pi/18$, respectively. Here, we fix $\sigma_{s\varphi}=\pi/36$, and the viewing azimuth coincides with $\varphi_{\textrm{s}}$. All other parameters are the same as in Fig. \ref{fig8}.}      
\label{fig9}                 
\end{figure*}

Immediately following, we examine the cross-section of the spot in the vertical plane through Fig. \ref{fig9}. From the six panels, we confirm that the spot emission is symmetric about the equatorial plane, regardless of the parameter values. As expected, increasing $\sigma_{s\theta}$ extends the spot in the vertical direction. From panel (d), we find that when $\sigma_{s\theta}$ is large but $\sigma_{sr}$ is small, the localized spot resembles the cross-section of a shock wave.

In total, our accretion model contains $17$ structural parameters. Specifically, $r_{\textrm{in}}$ sets the inner boundary of the accretion model; $p_{1}$ and $p_{2}$ control the overall radial decay of the disk emission; $\sigma_{d\theta}$ and $\beta$ characterize the disk thickness and flaring. The parameters $j_{\textrm{p}}$, $w_{\textrm{p}}$, and $r_{\textrm{p}}$ introduce a plateau that phenomenologically models a local enhancement or compensation of radiative efficiency. The parameters $r_{\textrm{b}}$, $\sigma_{br}$, and $\sigma_{b\theta}$ place a Gaussian bump, which can mimic outwardly propagating shock waves triggered by accretion rate fluctuations, MRI driven turbulence, or disk-jet interactions. The parameters $(r_{\textrm{s}},\theta_{\textrm{s}},\varphi_{\textrm{s}})$, $\sigma_{sr}$, $\sigma_{s\theta}$, and $\sigma_{s\varphi}$ introduce localized spots that can represent compact flaring regions caused by TDE, stream-stream collisions, or magnetic reconnection episodes. Owing to this multi-parameter construction, our model offers a high degree of flexibility to qualitatively capture a wide variety of potential high-energy events around black holes, despite not being tied to a specific dynamical simulation.
\subsubsection{Dynamic profiles}
\begin{figure*}
\centering                   
\includegraphics[width=14cm]{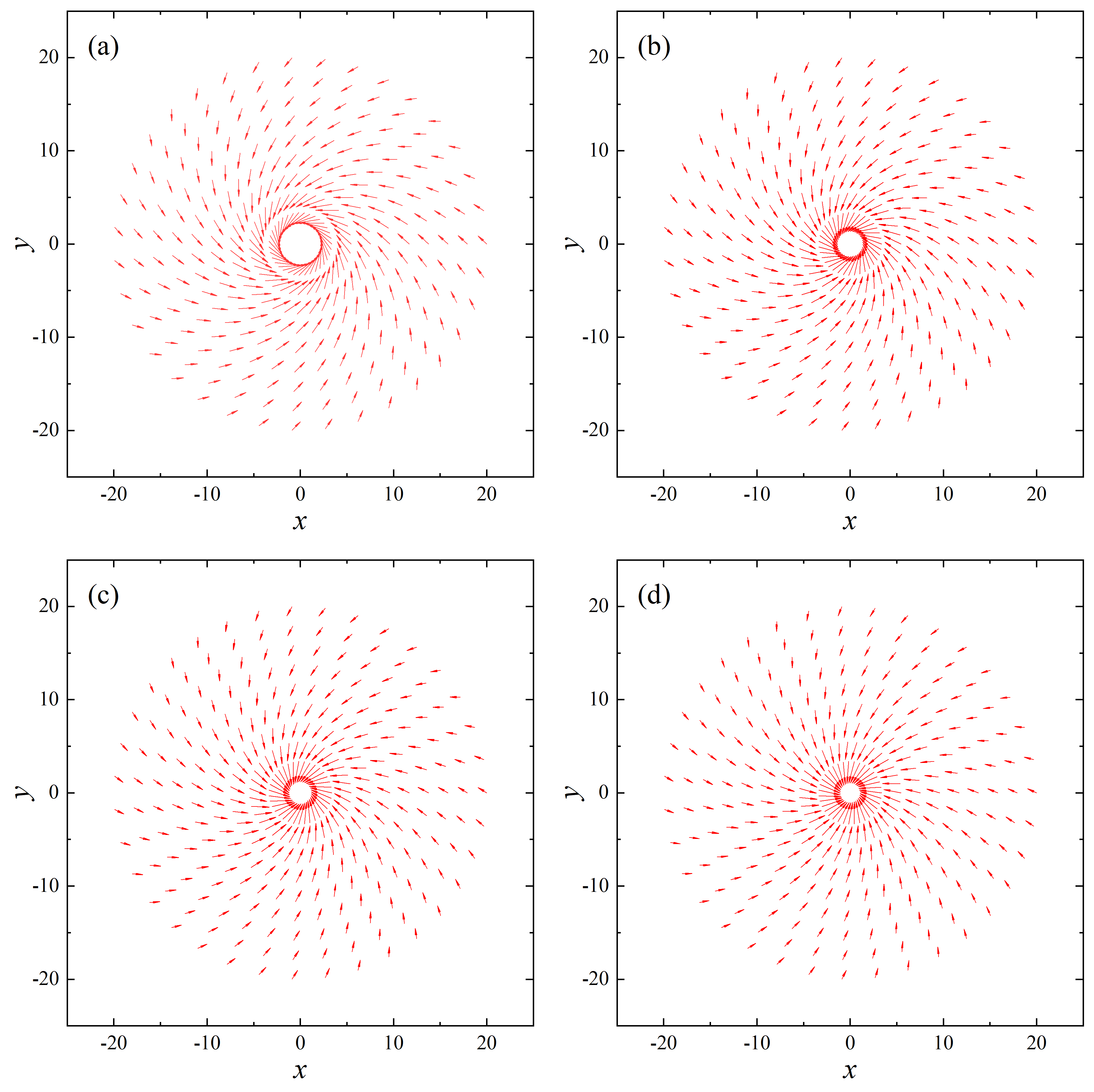}
\caption{Velocity distributions of the accretion flow in the equatorial plane for different parameter values. From panel (a) to (d), $\lambda=0.01$, $5$, $10$, and $20$, respectively. We fix $V_{\textrm{max}}=\psi=0.9$ and $p_{3}=0.5$.}      
\label{fig10}                 
\end{figure*}
For the dynamical behavior of the accretion flow, we adopt a global rotation approximation and neglect the dynamical differences among individual components. This approximation is analogous to the angular momentum transport mechanism in astrophysical accretion disks. Fig. \ref{fig10} shows the influence of different $\lambda$ values on the velocity field of the accretion flow in the equatorial plane. We find that the velocity of the accretion flow is relatively low at large radii and increases as $r$ decreases. When $\lambda$ is small, the azimuthal velocity of the accreting material is clearly visible from the outer to the inner region. However, as $\lambda$ increases, the velocity near the center becomes increasingly dominated by the radial component, as is evident in panel (d). Furthermore, larger values of $\lambda$ can also affect the velocity at larger radii, causing the radially dominated accretion to occur at an earlier stage.
\subsection{Black hole images}
\subsubsection{Ray-tracing and radiative transfer}
Before presenting the simulated black hole images, it is useful to briefly review the phenomenological procedure for producing such images. Given the line element of the Kerr spacetime, we can define the Lagrangian governing photon propagation as
\begin{eqnarray}\label{28}
\mathscr{L} = \frac{1}{2}g_{\mu\nu}\dot{x}^{\mu}\dot{x}^{\nu},
\end{eqnarray}
where $\dot{x}^{\mu}=\textrm{d}x^{\mu}/\textrm{d}\tau$ is the photon four-velocity and $\tau$ is the affine parameter. From the Euler-Lagrange equations, the canonical momentum is defined as $p_{\mu}=\partial \mathscr{L} / \partial \dot{x}^{\mu}$. Since the Kerr metric does not depend explicitly on the coordinates $t$ and $\varphi$, both $-p_{t}$ and $p_{\varphi}$ are conserved quantities, corresponding to the specific energy $E$ and specific angular momentum $L$ of the photon, respectively.

By means of the Legendre transformation, $\mathscr{H}= p_{\mu}\dot{x}^{\mu}-\mathscr{L}$, we obtain the Hamiltonian describing the photon motion,
\begin{equation}\label{29}
\mathscr{H} = \frac{1}{2}g^{\mu\nu}p_{\mu}p_{\nu},
\end{equation}
where $g^{\mu\nu}$ is the contravariant metric. From the covariant metric given in Eq. \eqref{25}, the nonvanishing components of $g^{\mu\nu}$ are \cite{Wu:2021rrd}:
\begin{eqnarray}
g^{tt} &=& -\frac{\left(r^{2}+a^{2}\right)^{2}-\Delta a^{2}\sin^{2}\theta}{\Delta\Sigma}, \label{30} \\
g^{rr} &=& \frac{\Delta}{\Sigma}, \label{31} \\
g^{\theta\theta} &=& \frac{1}{\Sigma}, \label{32} \\
g^{\varphi\varphi} &=& \frac{\Sigma-2r}{\Delta\Sigma\sin^{2}\theta}, \label{33} \\
g^{t\varphi} &=& g^{\varphi t} = -\frac{2ar}{\Delta\Sigma}. \label{34}
\end{eqnarray}
With the initial coordinates $(t,r,\theta,\varphi)$ and canonical momenta $(p_{t},p_{r},p_{\theta},p_{\varphi})$ of a photon, its trajectory can be determined by solving Hamilton's equations:
\begin{equation}\label{35}
\dot{x}^{\mu}=\frac{\partial \mathscr{H}}{\partial p_{\mu}}, \quad \dot{p}_{\mu} = -\frac{\partial \mathscr{H}}{\partial x^{\mu}}.
\end{equation}

Next, we obtain the initial conditions of the light rays in the local coordinates of the black hole following the method described in ODYSSEY \cite{Pu:2016eml}. We assume that the observer's local reference frame is denoted by $xyz$, while the local coordinates of the black hole are denoted by $x^{\prime}y^{\prime}z^{\prime}$. Here, the $oz$ axis points from the observer toward the origin $o^{\prime}$ of the black hole frame. The angle between $o^{\prime}o$ and the black hole spin axis, i.e., the $z^{\prime}$-axis, is defined as the observation inclination angle, denoted by $\Theta$; the angle between the projection of $o^{\prime}o$ onto the equatorial plane $\overline{x^{\prime}o^{\prime}y^{\prime}}$ and the $x^{\prime}$-axis is defined as the observation position angle, denoted by $\Phi$. We choose the $\overline{xoy}$ plane of the observer's local frame as the image plane, where each point is treated as the starting pixel of a light ray. The initial coordinates of a light ray in the observer's frame are written as $(x,y,0)$. The transformation between $(x,y,0)$ and the black hole coordinates $x^{\prime}y^{\prime}z^{\prime}$ is given by
\begin{eqnarray}
x^{\prime} &=& \mathscr{T}\cos\Phi - x\sin\Phi, \label{36} \\
y^{\prime} &=& \mathscr{T}\sin\Phi + x\cos\Phi, \label{37} \\
z^{\prime} &=& \left(r_{\textrm{obs}} - z\right)\cos\Theta + y\sin\Theta, \label{38}
\end{eqnarray}
where $r_{\textrm{obs}}$ denotes the observer distance, i.e., the length of $oo^{\prime}$, and the function $\mathscr{T}$ takes the form
\begin{equation}\label{39}
\mathscr{T} = \left(\sqrt{r_{\textrm{obs}}^{2}+a^{2}}-z\right)\sin\Theta-y\cos\Theta.
\end{equation}
Furthermore, the initial position of the light ray can be mapped into the black hole coordinates $(t,r,\theta,\varphi)$ via\footnote{This transformation is specifically tailored for mapping from the auxiliary Cartesian coordinates to the Boyer-Lindquist coordinates of the Kerr spacetime. Applying the same procedure to other asymptotically flat axisymmetric spacetimes may introduce systematic offsets. Nevertheless, for sufficiently large observer distances $r_{\textrm{obs}}$, such offsets become negligibly small, and the transformation can be approximately treated as a universal method.}
\begin{eqnarray}
t &=& 0, \label{40} \\
r &=& \sqrt{\frac{\mathscr{D}+\sqrt{\mathscr{D}^{2}+4a^{2}z^{\prime2}}}{2}}, \label{41} \\
\theta &=& \arccos\left(\frac{z^{\prime}}{r}\right), \label{42} \\
\varphi &=& \textrm{atan}2\left(y^{\prime},x^{\prime}\right), \label{43}
\end{eqnarray}
where $\mathscr{D}$ is given by
\begin{equation}\label{44}
\mathscr{D} = x^{\prime2} + y^{\prime2} + z^{\prime2} - a^{2}.
\end{equation}

We fix the observer distance to $r_{\textrm{obs}}=1000$, for which the local spacetime of the observer can be approximated as flat. In this case, the light rays can be traced backward perpendicular to the image plane. Specifically, the initial three-velocity of a light ray measured by the observer is $(\dot{x},\dot{y},\dot{z})=(0,0,-1)$. Substituting this condition into the differentiated forms of Eqs. \eqref{36}--\eqref{38}, we obtain the corresponding three-velocity in the $x^{\prime}y^{\prime}z^{\prime}$ frame:
\begin{eqnarray}
\dot{x}^{\prime} &=& \sin\Theta\cos\Phi, \label{45} \\
\dot{y}^{\prime} &=& \sin\Theta\sin\Phi, \label{46} \\
\dot{z}^{\prime} &=& \cos\Theta. \label{47}
\end{eqnarray}
Similarly, substituting $(\dot{x}^{\prime},\dot{y}^{\prime},\dot{z}^{\prime})$ into the differentiated forms of Eqs. \eqref{41}--\eqref{43} yields $(\dot{r},\dot{\theta},\dot{\varphi})$ as
\begin{eqnarray}
\dot{r} &=& \frac{r\mathscr{R}\sin\theta\sin\Theta\cos\Psi+\mathscr{R}^{2}\cos\theta\cos\Theta}{\Sigma}, \label{48} \\
\dot{\theta} &=& \frac{\mathscr{R}\cos\theta\sin\Theta\cos\Psi-r\sin\theta\cos\Theta}{\Sigma}, \label{49} \\
\dot{\varphi} &=& -\frac{\sin\Theta\sin\Psi}{\mathscr{R}\sin\theta}, \label{50}
\end{eqnarray}
where $\mathscr{R}=\sqrt{r^{2}+a^{2}}$ and $\Psi=\varphi-\Phi$. Once the three-velocity components are obtained, the corresponding radial and polar canonical momenta, $p_{r}$ and $p_{\theta}$, can also be derived. 

Next, from
\begin{equation}\label{51}
p_{t} = g_{tt}\dot{t} + g_{t\varphi}\dot{\varphi} = -\left(1-\frac{2r}{\Sigma}\right)\dot{t}-\frac{2ar\sin^{2}\theta}{\Sigma}\dot{\varphi}, 
\end{equation}
we solve for
\begin{equation}\label{52}
\dot{t} =\left(\frac{\Sigma}{\Sigma-2r}\right)\left(-p_{t} - \frac{2ar\sin^{2}\theta}{\Sigma}\dot{\varphi}\right).
\end{equation}
Substituting this into the Lagrangian and imposing the null condition, we obtain the time component of the canonical momentum:
\begin{equation}\label{53}
(-p_{t})^{2} = \left(\frac{\Sigma-2r}{\Sigma\Delta}\right)\left(\Sigma\dot{r}^{2}+\Sigma\Delta\dot{\theta}^{2}\right)+\Delta\dot{\varphi}^{2}\sin^{2}\theta.
\end{equation}

Finally, using the Hamiltonian constraint $\mathscr{H}=0$, we further obtain the azimuthal canonical momentum $p_{\varphi}$. It is worth noting that all canonical momenta can be renormalized by $p_{t}$ \cite{Cunha:2016bpi}. At this point, for each pixel in the image plane, the corresponding initial conditions can be obtained, and the trajectory of each light ray can be determined by integrating the geodesic equations using a fifth- and sixth-order Runge-Kutta-Fehlberg (RKF56) integrator with an adaptive step size. 

When a light ray does not intersect the accretion medium and falls directly into the black hole, the corresponding pixel is assigned a specific intensity of zero. When a light ray propagates through the accretion medium, we need to compute the specific intensity it carries by solving the radiative transfer equation. The covariant radiative transfer equation is written as \cite{Pu:2016eml}
\begin{equation}\label{54}
\frac{d\mathscr{I}}{\textrm{d}\tau} = \mathscr{J}-\mathscr{A}\mathscr{I},
\end{equation}
where $\mathscr{I}=I_{\nu}/\nu^{3}$, $\mathscr{A}=\nu\alpha_{\nu}$, and $\mathscr{J}=j_{\nu}/\nu^{2}$ are the invariants of the specific intensity, absorption coefficient, and emission coefficient, respectively, with quantities subscripted by ``$\nu$'' being those measured in the local frame. Equivalently, we have
\begin{equation}\label{55}
\frac{\textrm{d}}{\textrm{d}\tau}\left(\frac{I_{\nu}}{\nu^{3}}\right) = \frac{j_{\nu}}{\nu^{2}} - \nu\alpha_{\nu}\left(\frac{I_{\nu}}{\nu^{3}}\right).
\end{equation}
It should be emphasized that our accretion model does not focus on features at any specific frequency, but rather on the overall geometric and optical depth distribution. Therefore, the emission frequency can be decoupled from the radiative transfer by introducing a redshift factor. The redshift factor $g$ is defined as
\begin{equation}\label{56}
g=\frac{\nu_{\textrm{obs}}}{\nu}=\frac{\left(p_{\mu}u^{\mu}\right)_{\textrm{obs}}}{\left(p_{\mu}u^{\mu}\right)_{\textrm{source}}},
\end{equation}
where $p_{\mu}$ is the photon canonical momentum, and $u^{\mu}$ is the four-velocity of the observer or the emission source, depending on the subscript. Substituting $\nu=\nu_{\textrm{obs}}/g$ into Eq. \eqref{55} yields
\begin{equation}\label{57}
\frac{\textrm{d}}{\textrm{d}\tau}\left(g^{3}\frac{I_{\nu}}{\nu_{\textrm{obs}}^{3}}\right) = g^{2}\frac{j_{\nu}}{\nu_{\textrm{obs}}^{2}} - \alpha_{\nu}\frac{\nu_{\textrm{obs}}}{g}\left(g^{3}\frac{I_{\nu}}{\nu_{\textrm{obs}}^{3}}\right).
\end{equation}

Since the photon energy has been renormalized, we can safely set $\nu_{\textrm{obs}}=1$ without affecting the radiative transfer calculation. Thus, we have 
\begin{equation}\label{58}
\frac{\textrm{d}}{\textrm{d}\tau}\left(g^{3}I_{\nu}\right) = g^{2}j_{\nu}-\frac{\alpha_{\nu}}{g}\left(g^{3}I_{\nu}\right).
\end{equation}
Note that $g^{3}I_{\nu}$ is precisely the invariant $\mathscr{I}$, which is the specific intensity of the light ray associated with each pixel. In what follows, we denote it by $I_{\textrm{obs}}$. Following \cite{Pu:2016eml}, we further write the differential equations for solving $I_{\textrm{obs}}$ as 
\begin{eqnarray}
\frac{\textrm{d}\varrho}{\textrm{d}\tau}=\frac{\alpha_{\nu}}{g}, \label{59} \\
\frac{\textrm{d}I_{\textrm{obs}}}{\textrm{d}\tau}=g^{2}j_{\nu}\textrm{e}^{-\varrho}, \label{60}
\end{eqnarray}
where $\varrho$ is the optical depth. By supplementing Hamilton's equations \eqref{35} with Eqs. \eqref{59} and \eqref{60}, we can simultaneously solve the ray-tracing and the radiative transfer.
\subsubsection{Image features}
\begin{figure*}
\centering                   
\includegraphics[width=15cm]{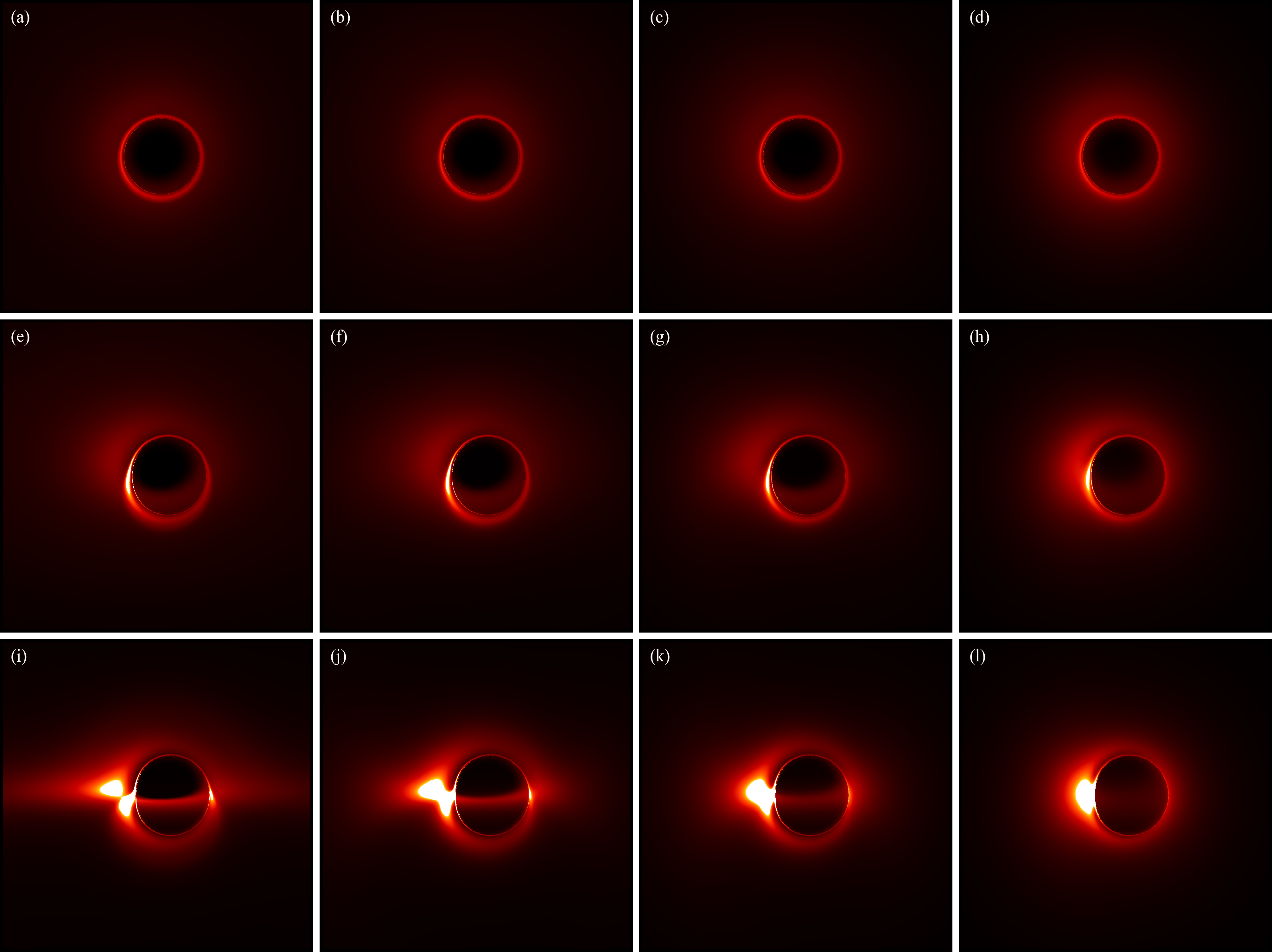}
\caption{Kerr black hole images for different observation inclination angles $\Theta$ and disk thickness parameters $\sigma_{d\theta}$. From left to right, $\sigma_{d\theta}=0.01$, $0.05$, $0.1$, and $0.25$; from top to bottom, the observation inclinations are $17^{\circ}$, $50^{\circ}$, and $80^{\circ}$, respectively. We fix $p_{1}=-1.5$, $p_{2}=-0.5$, and $\beta=0.1$; the plateau is placed at $r_{\textrm{p}}=1.5$, with $w_{\textrm{p}}=3$ and $j_{\textrm{p}}=1$. For visualization, the maximum values of the color bars are set to $0.2$, $1.2$, $2$, and $4$ from left to right, with a fixed gamma value of $0.7$.}      
\label{fig11}                 
\end{figure*}

We fix the field of view to $x$, $y$ $\in$ $[-20,20]$ M, with a resolution of $1000 \times 1000$ pixels. The accretion material extends down to the event horizon, and the dynamical parameters are fixed to $V_{\textrm{max}}=\psi=0.9$, $p_{3}=0.5$, and $\lambda=10$. Fig. \ref{fig11} shows the Kerr black hole images for different observation inclination angles and disk thickness parameter, with the spin parameter fixed to $a=0.94$, under the emission model $j_{\nu}(r,\theta,\varphi)=j_{\textrm{d}}(r,\theta)$.

In each panel, a thin and sharp critical curve can be clearly identified, whose shape is determined by the inclination angle and the black hole spin. At low inclinations, the critical curve is nearly circular, while at higher inclinations it transitions to a characteristic ``D''-shaped morphology. On the left side of the critical curve, a prominent bright arc (or bright spot) is attached, which arises from the combined effects of Doppler boosting and the plateau emission. As a result, this feature becomes particularly pronounced at high inclination angles.

The influence of the disk thickness parameter on the image morphology is twofold. First, increasing $\sigma_{d\theta}$ enhances the overall brightness of the image. This can be inferred from the fact that, from left to right, the visual brightness distribution appears roughly consistent across panels while the color bar ranges increasing progressively. Second, the black hole shadow inside the critical curve is sensitive to the disk thickness. At $\Theta=17^{\circ}$, the critical curve does not coincide with the shadow boundary; increasing $\sigma_{d\theta}$ slightly reduces the apparent shadow size. This shadow shrinking effect becomes more evident at $\Theta=50^{\circ}$, where the thickened accretion environment can intercept photons that would otherwise fall directly into the black hole. When the inclination reaches $80^{\circ}$, the overall boundary of the shadow approaches the critical curve, but is divided into upper and lower parts by a sharply defined bright arc, as seen in panel (i). As $\sigma_{d\theta}$ increases, this bright arc becomes thicker and more diffuse. At $\sigma_{d\theta}=0.25$, the arc becomes so diffuse that it is barely discernible, and the shadow is once again displayed as a single region whose boundary approaches the critical curve, as shown in panel (l). These results indicate that when the inner boundary of the accretion disk intersects the event horizon and the disk is nearly geometrically thin, the observable shadow boundary generally deviates from the critical curve, manifesting as an inner shadow \cite{2021ApJ...918....6C}. When the geometric thickness of the disk becomes non-negligible, the observable shadow boundary approaches the critical curve, a trend that has also been confirmed in \cite{2021ApJ...920..155B}.

\begin{figure*}
\centering                   
\includegraphics[width=14cm]{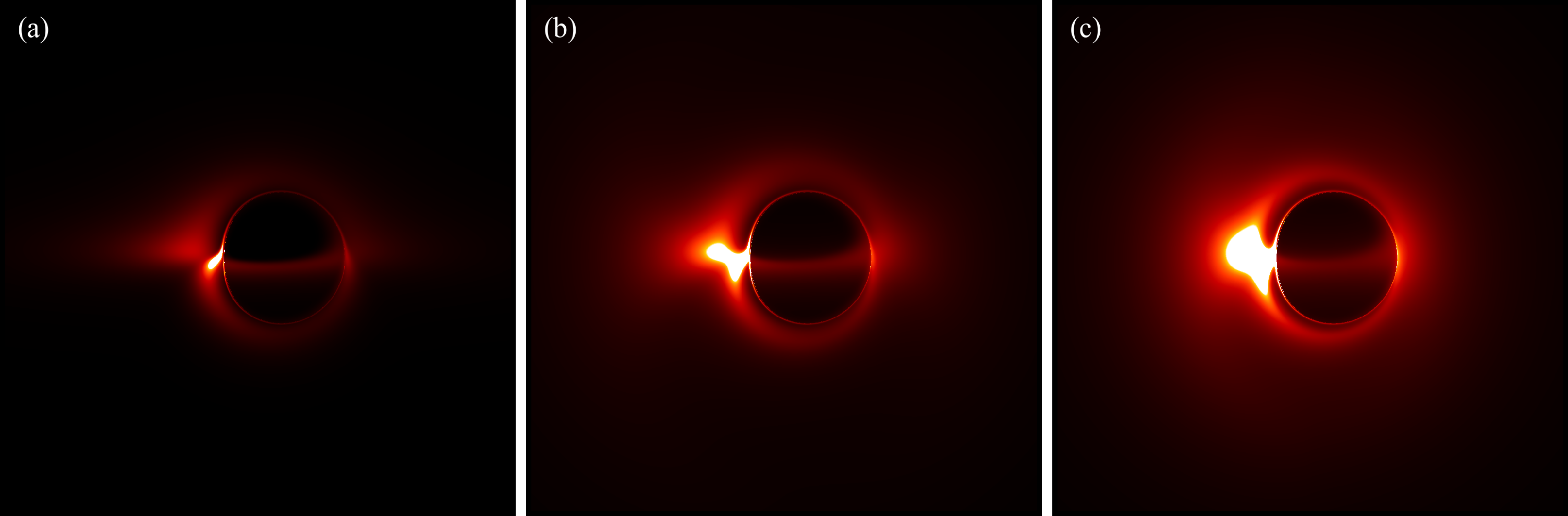}
\caption{Influence of increasing $\beta$ on the image morphology of Kerr black holes. From left to right, $\beta=0.01$, $0.15$, and $0.25$, respectively. Here, we fix $\sigma_{d\theta}=0.05$ and the observation angle $\Theta=80^{\circ}$. In each panel, the color bar ranges from $0$ to $2$, with a gamma value of $0.7$.}      
\label{fig12}                 
\end{figure*}
We fix the observation angle to $80^{\circ}$ and $\sigma_{d\theta}=0.05$ to examine the influence of $\beta$ on the image features, as shown in Fig. \ref{fig12}. We find that increasing $\beta$ enhances the overall brightness of the image while keeping the critical curve clearly visible. However, the blurring effect of $\beta$ on the bright arc that divides the shadow is less pronounced than that of $\sigma_{d\theta}$. Furthermore, combining Figs. \ref{fig11} and \ref{fig12}, we observe that at high inclination angles, the images of geometrically thin (or nearly so) disks exhibit a mushroom-like morphology, where the direct image contributes to the cap of the mushroom, while the secondary and higher-order images form the stalk. For geometrically thick disks, the clear mushroom shape is broken, and is instead replaced by a bright ring with attached spots that enclose the shadow.

\begin{figure*}
\centering                   
\includegraphics[width=15cm]{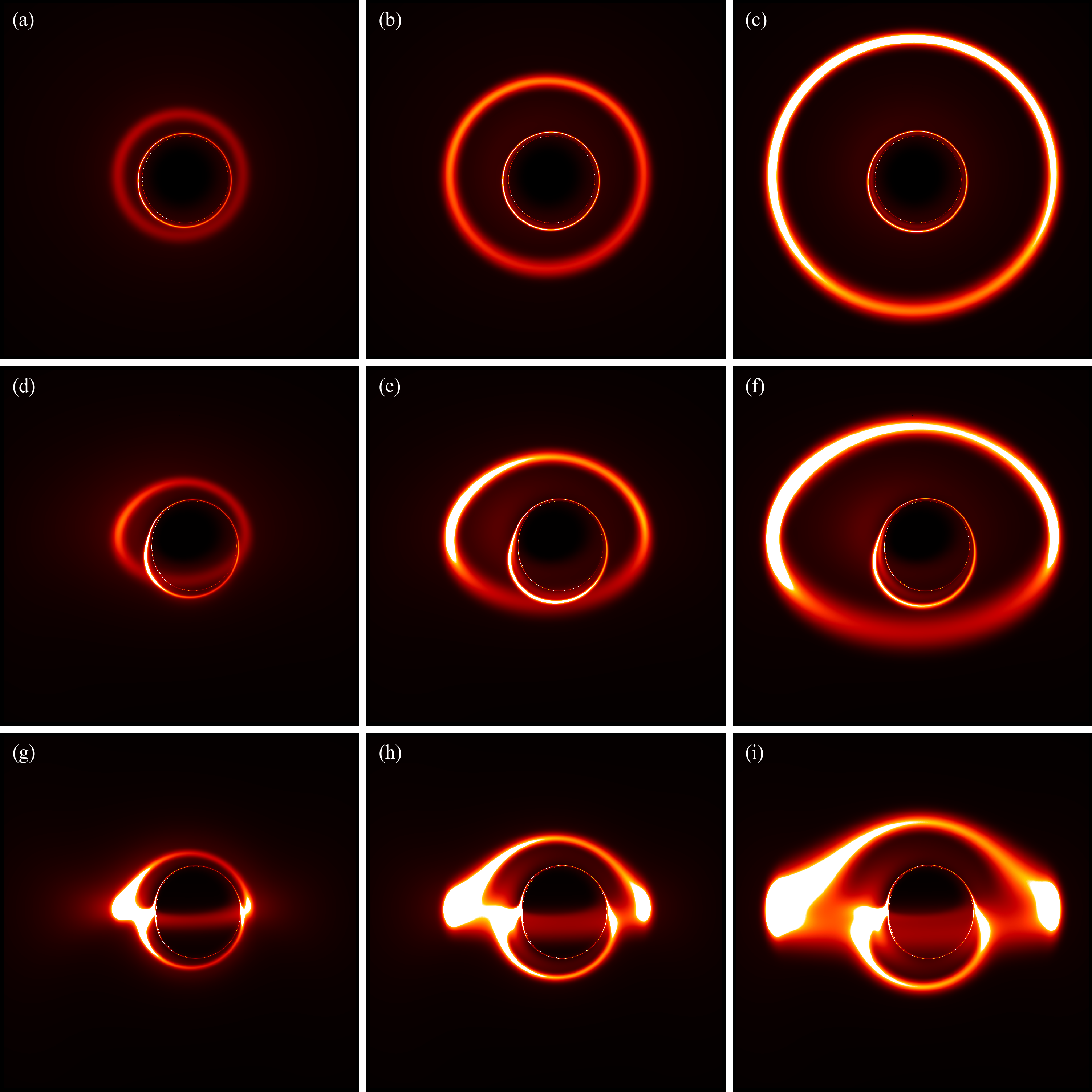}
\caption{Kerr black hole images after introducing a Gaussian bump at different radial positions. From top to bottom, the observation inclinations are $17^{\circ}$, $50^{\circ}$, and $80^{\circ}$, respectively; from left to right, the bump position $r_{\textrm{b}}=6$, $10$, and $15$, respectively. We fix $p_{1}=-1.5$, $p_{2}=-0.5$, $\sigma_{d\theta}=0.05$, $\beta=0.1$, $r_{\textrm{p}}=1.5$, and $w_{\textrm{p}}=3$; the bump morphology parameters are set to $\sigma_{br}=0.5$ and $\sigma_{b\theta}=0.1$. The color bar is fixed to the range $[0,2]$, with a gamma value of $0.7$.}      
\label{fig13}                 
\end{figure*}
Next, Fig. \ref{fig13} shows the images after introducing a Gaussian bump at different radial positions within the disk, with $\sigma_{d\theta}=0.05$, $\beta=0.1$, and $\sigma_{br}=0.5$ fixed, while all other parameters are the same as in Fig. \ref{fig11}. When the observation inclination is $\Theta=17^{\circ}$ (first row), the contribution of $j_{\textrm{b}}(r,\theta)$ mainly appears as a prominent brightness asymmetry in the ring, which moves outward and becomes gradually brighter as $r_{\textrm{b}}$ increases. Increasing $\Theta$ to $50^{\circ}$, this bright ring is compressed in the vertical direction, taking on a pebble-like shape. As the inclination increases further, the ring is deformed by gravitational lensing into a cap-like structure, accompanied by the appearance of multiple bright spots in the image. Importantly, the critical curve remains visible regardless of the bump position.
\begin{figure*}
\centering                   
\includegraphics[width=15cm]{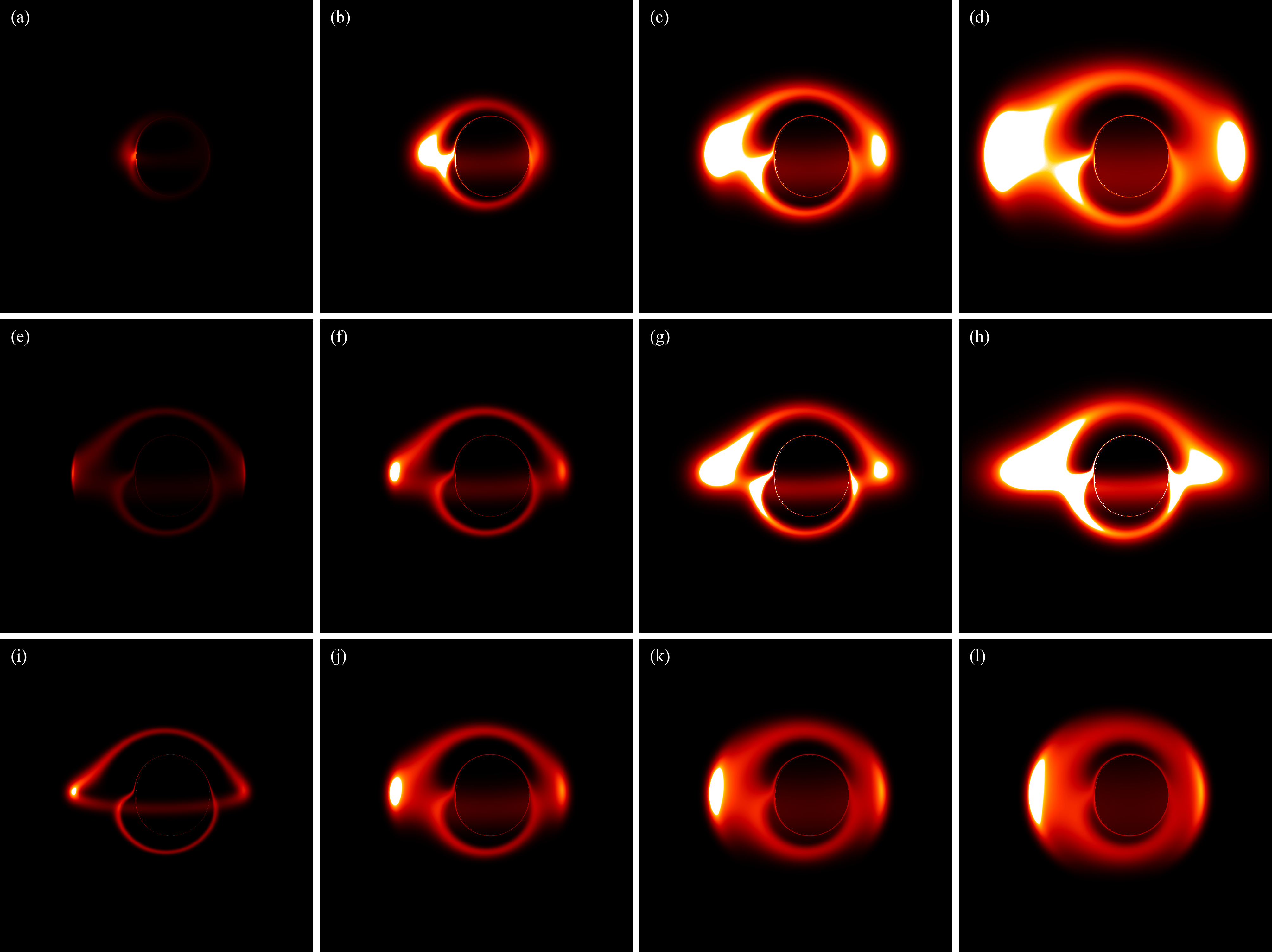}
\caption{Kerr black hole images as functions of $r_{\textrm{b}}$, $\sigma_{br}$, and $\sigma_{b\theta}$. The observation inclination is fixed to $80^{\circ}$, and the disk contribution $j_{\textrm{d}}$ is turned off. In the first row, we fix $\sigma_{br}=1$ and $\sigma_{b\theta}=0.2$, with $r_{\textrm{b}}=2$, $6$, $10$, and $15$ from left to right. In the second row, we fix $r_{\textrm{b}}=10$ and $\sigma_{b\theta}=0.1$, with $\sigma_{br}=0.1$, $0.5$, $1.5$, and $3$ from left to right. In the last row, we fix $r_{\textrm{b}}=10$ and $\sigma_{br}=0.5$, with $\sigma_{b\theta}=0.05$, $0.15$, $0.25$, and $0.35$ from left to right. The color bar is fixed to the range $[0,5]$, with a gamma value of $0.7$.}      
\label{fig14}                 
\end{figure*}

Thanks to the modular nature of our accretion environment, we can isolate the Gaussian bump component and study the image features with the disk contribution turned off. In this case, the black hole is effectively surrounded by a torus-like structure. Fig. \ref{fig14} presents the resulting images, showing the influence of $r_{\textrm{b}}$ (first row), $\sigma_{br}$ (second row), and $\sigma_{b\theta}$ (third row), respectively. We find that increasing $r_{\textrm{b}}$ not only shifts the bright ring outward but also significantly enhances its brightness and introduces several additional bright spots in the image. Moreover, since the Gaussian bump does not extend down to the event horizon, a semicircular shadow-like region can be identified above and below the critical curve, in addition to the shadow inside the critical curve. The influence of $\sigma_{br}$ on the image features is mainly limited to modifying the brightness and refining morphological details. For example, in panel (e), a faint mushroom-like pattern can be discerned, with eyebrow-shaped bright streaks visible at both ends. When $\sigma_{br}$ is increased by a factor of $30$, as shown in panel (h), the overall mushroom structure does not change significantly, but the brightness in several regions is notably enhanced, and the eyebrow-shaped streaks become smoother. When we fix $r_{\textrm{b}}=10$ and $\sigma_{br}=0.5$ and increase $\sigma_{b\theta}$, the overall brightness remains nearly unchanged, except for the expansion of the spot size, while the morphology undergoes a qualitative transformation. This is because $\sigma_{b\theta}$ controls the vertical extent of the Gaussian bump; a larger $\sigma_{b\theta}$ indicates a thicker torus, which necessarily leads to a vertical extension of the image structure.

\begin{figure*}
\centering                   
\includegraphics[width=15cm]{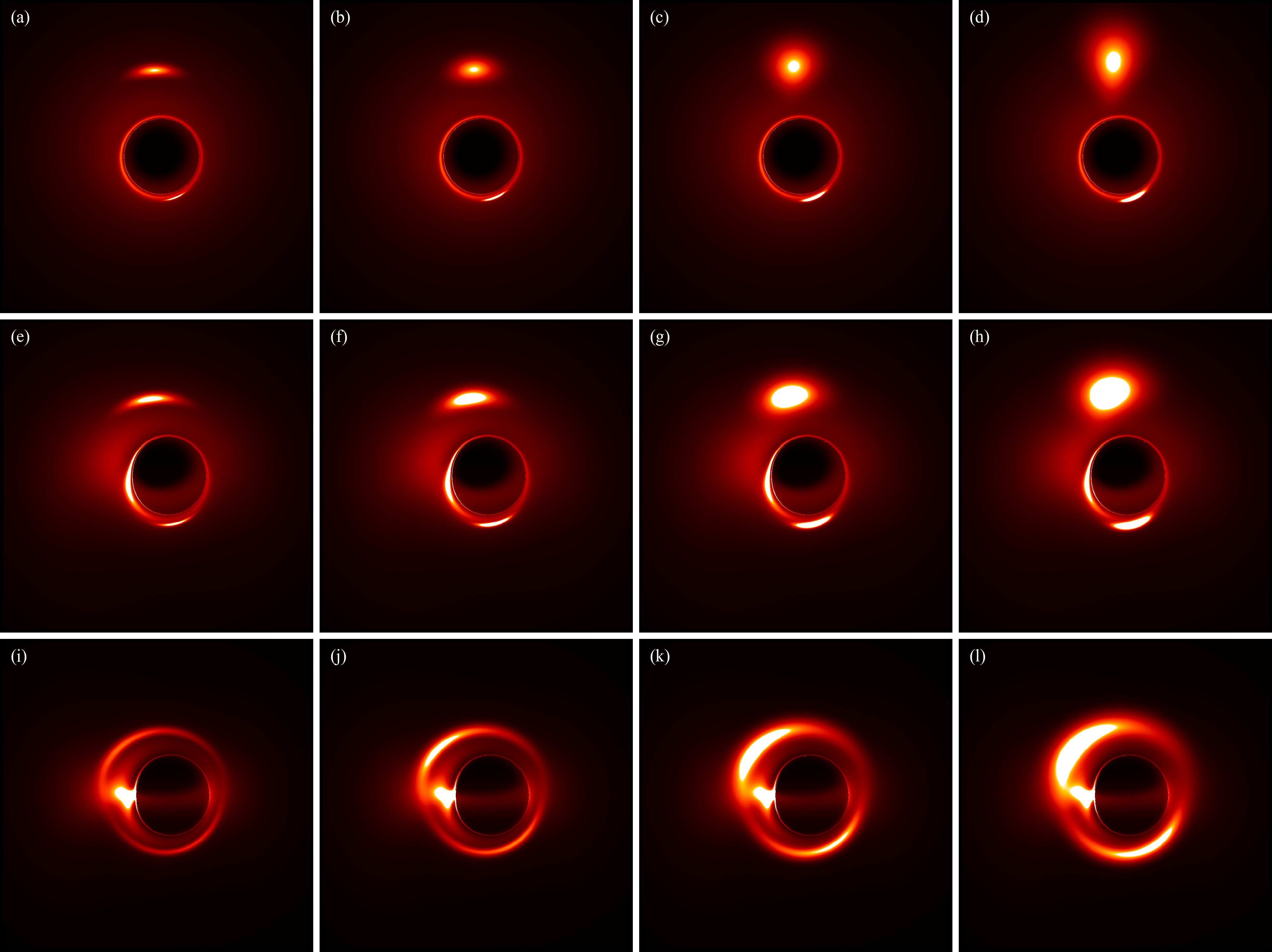}
\caption{Kerr black hole images after introducing a localized spot at $(r_{\textrm{s}},\theta_{\textrm{s}},\varphi_{\textrm{s}})=(10,\pi/2,\pi)$, for different observation inclination angles and $\sigma_{sr}$ values. From top to bottom, the observation inclinations are $17^{\circ}$, $50^{\circ}$, and $80^{\circ}$, respectively; from left to right, $\sigma_{sr}=0.5$, $1$, $2$, and $3$, respectively. We fix the parameters associated with $j_{\textrm{d}}$ to $\sigma_{d\theta}=\beta=0.1$, $p_{1}=-1.5$, $p_{2}=-0.5$, $r_{\textrm{p}}=1.5$, $j_{\textrm{p}}=1$, and $w_{\textrm{p}}=3$; the parameters related to the localized spot are fixed to $\sigma_{s\theta}=\pi/36$ and $\sigma_{s\varphi}=\pi/30$. The color bar maxima are $1.5$ for the first and second rows, and $3$ for the third row.}      
\label{fig15}                 
\end{figure*}
Finally, we introduce a localized spot at $(r_{\textrm{s}},\theta_{\textrm{s}},\varphi_{\textrm{s}})=(10,\pi/2,\pi)$, with $\sigma_{s\varphi}=\pi/30$ and $\sigma_{s\theta}=\pi/36$, to examine the influence of the observation inclination and $\sigma_{sr}$ on the image features, as shown in Fig. \ref{fig15}. The results show that at low inclination angles, the localized spot appears as a compact bright spot in the image. Its direct image is prominent, while the secondary and higher-order images, formed by gravitational lensing, appear as crescent-shaped features attached to the critical curve. When $\sigma_{sr}$ is small, the direct image of the spot is also crescent-like; however, as $\sigma_{sr}$ increases, the direct image stretches in the radial direction, taking on a teardrop shape, while the secondary images become slightly thicker and extend toward both ends. When the inclination is increased to $50^{\circ}$, the image of the localized spot becomes brighter and larger in extent. Meanwhile, the Doppler effect arising from the motion of the accretion flow provides additional bright spots on the left side of the critical curve, resulting in a multi-spot feature in the black hole image. When the observation approaches the edge-on configuration, the image of the localized spot is no longer a small bright patch, but is deformed by gravitational lensing into a ring with asymmetric brightness. This ring becomes thicker and brighter as $\sigma_{sr}$ increases.
\begin{figure*}
\centering                   
\includegraphics[width=15cm]{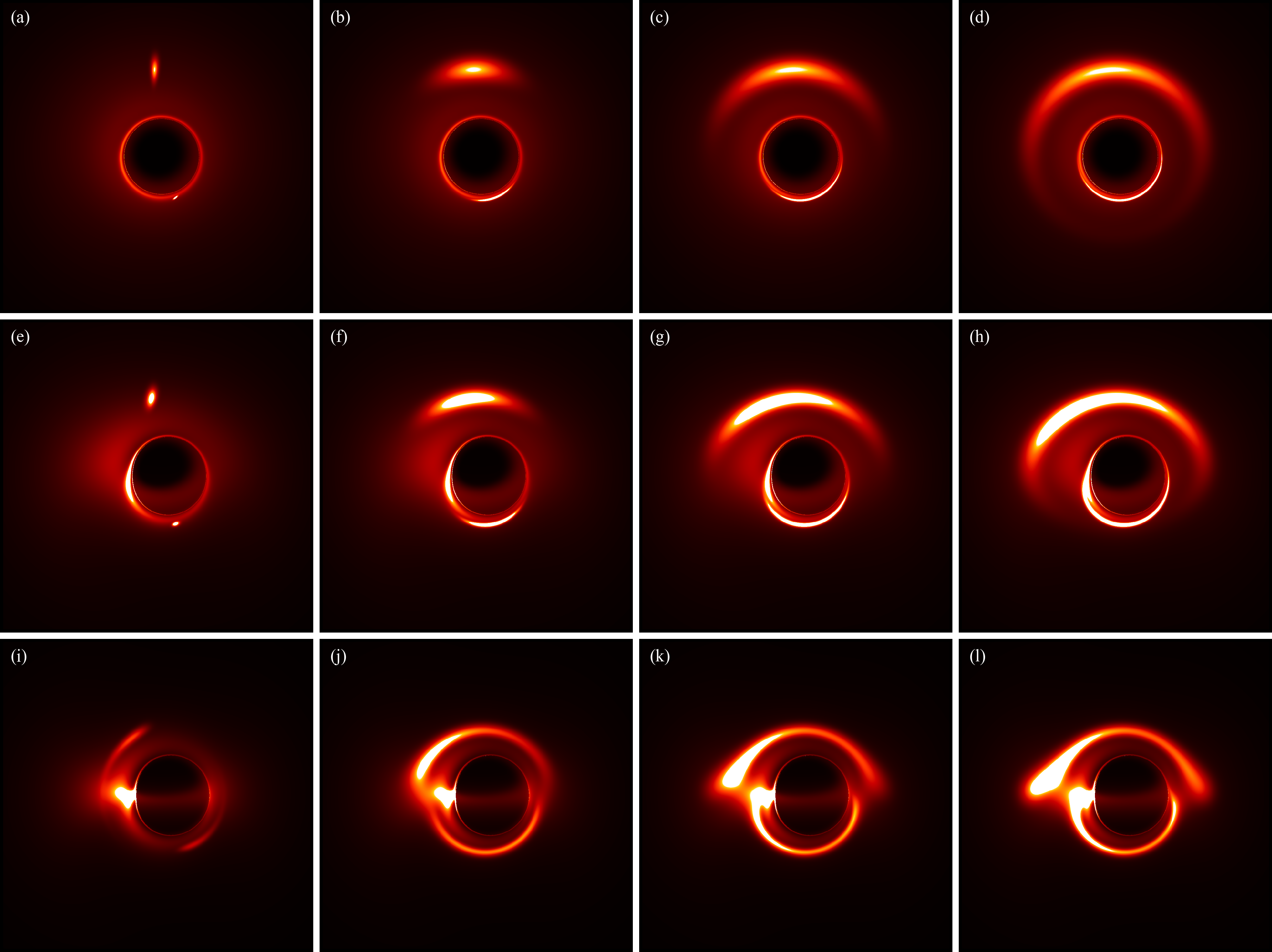}
\caption{Same as Fig. \ref{fig15}, but from left to right, $\sigma_{s\varphi}=1$, $10$, $20$, and $30$, respectively, with $\sigma_{sr}=1$ fixed.}      
\label{fig16}                 
\end{figure*}

It is expected that the image morphology of the localized spot depends not only on the inclination angle and $\sigma_{sr}$, but also on $\sigma_{s\varphi}$, which determines the azimuthal extent of the spot. We report this effect in Fig. \ref{fig16}. As shown in the first row, increasing $\sigma_{s\varphi}$ spreads the image of the localized spot along the azimuthal direction, an effect that applies to both the direct and secondary images.

\begin{figure*}
\centering                   
\includegraphics[width=15cm]{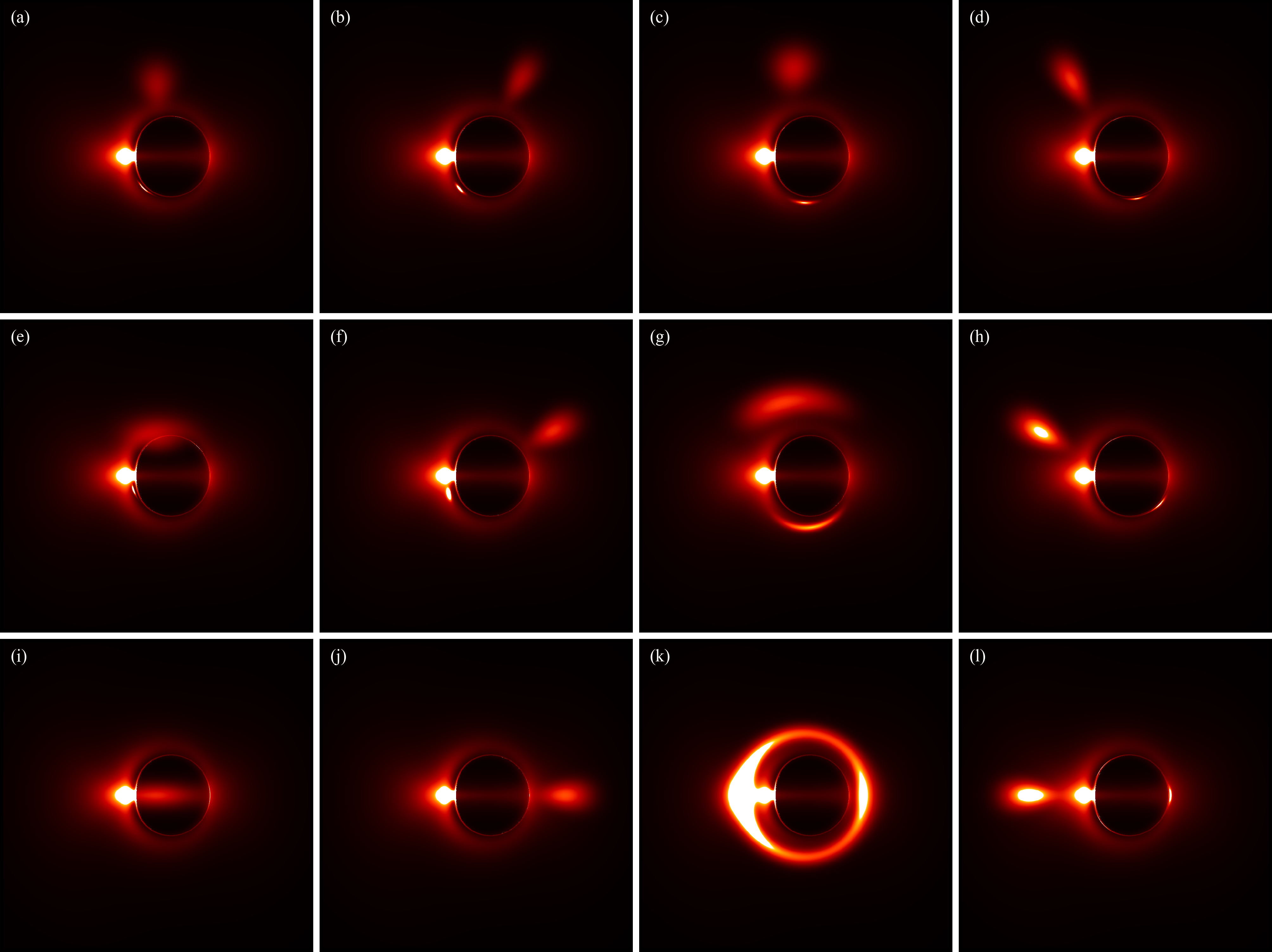}
\caption{Kerr black hole images for a localized spot placed at different positions on a sphere of radius $10$, observed at an inclination of $90^{\circ}$. From left to right, the azimuthal angles $\varphi_{\textrm{s}}$ of the spot are $0^{\circ}$, $90^{\circ}$, $180^{\circ}$, and $270^{\circ}$, respectively; from top to bottom, $\theta_{\textrm{s}}=30^{\circ}$, $60^{\circ}$, and $90^{\circ}$, respectively. We fix $\sigma_{s\varphi}=\pi/18$, $\sigma_{s\theta}=\pi/36$, and $\sigma_{sr}=2$. All other computational parameters are the same as in Fig. \ref{fig15}, and the color bar maximum is set to $4$ for all panels.}      
\label{fig17}                 
\end{figure*}

We fix $r_{\textrm{s}}=10$, $\sigma_{sr}=2$, $\sigma_{s\varphi}=\pi/18$, and $\sigma_{s\theta}=\pi/36$, and examine the images of the localized spot placed at different positions, as shown in Fig. \ref{fig17}. As $\varphi_{\textrm{s}}$ and $\theta_{\textrm{s}}$ vary, the position of the spot in the image changes accordingly, appearing in various forms such as compact spots or bright arcs in different regions outside the critical curve. There are also several interesting cases: the spot attaches to the critical curve, as shown in panel (e); appears inside the shadow, as displayed in panel (i); or forms an Einstein ring, as shown in panel (k).
\subsubsection{One-dimensional specific intensity}
It is worth noting that using the full image as the unit of inspection can only provide a rough inference of the accretion environment and the background spacetime parameters. A deeper inversion of the parameters, and even their constraints, requires additional measurements, such as the connection between the parameters and the one-dimensional intensity distribution of the image. In what follows, we examine, across different parameter spaces, the influence of individual components of the accretion environment on the intensity distribution along the $x$-axis of the observation screen, in order to qualitatively discuss the degeneracies among the model parameters and the potential observational prospects.
\begin{figure*}
\centering                   
\includegraphics[width=15cm]{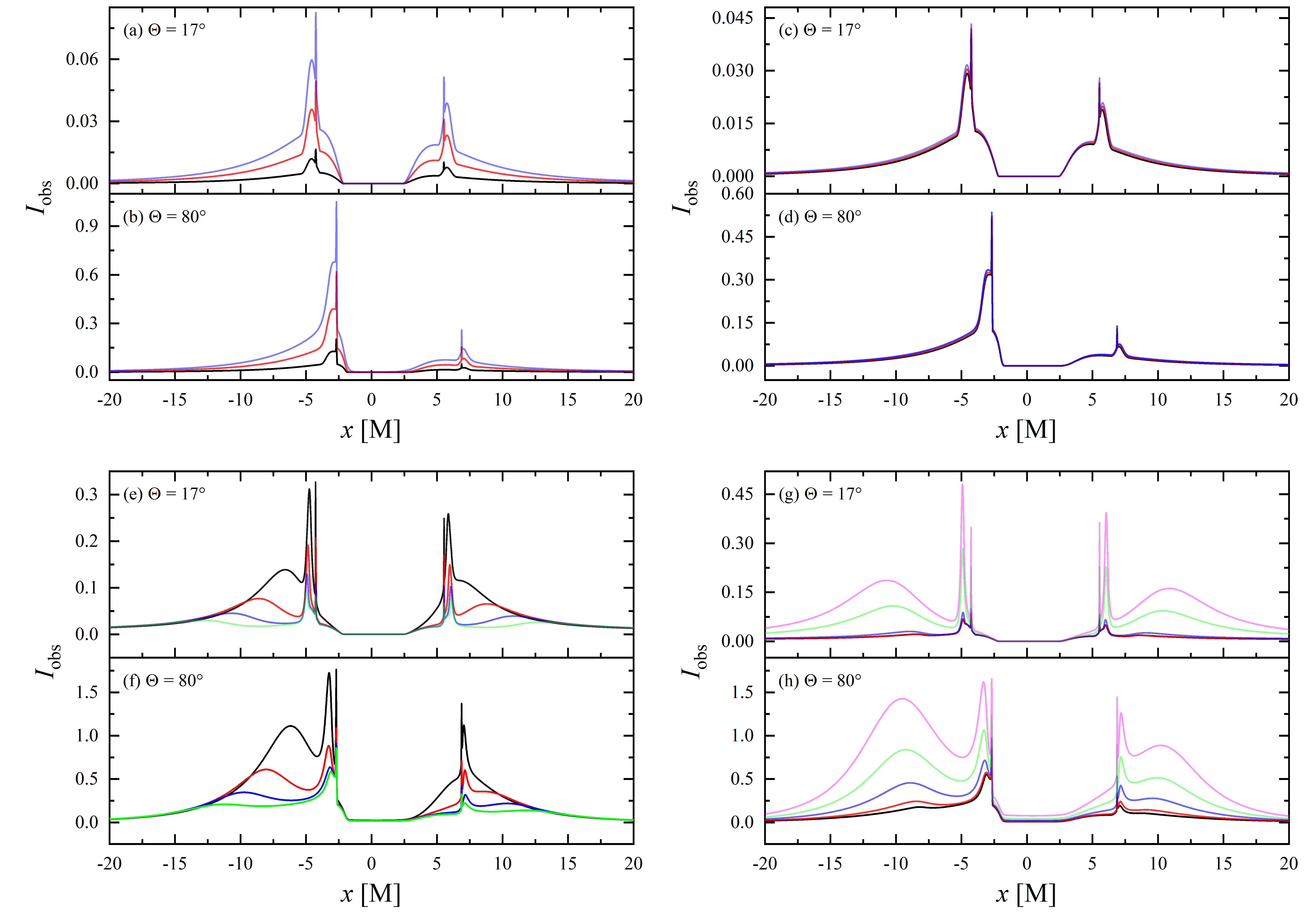}
\caption{Specific intensity distributions of light rays along the $x$-axis of the observation screen for different parameter spaces. The first row shows the results considering only the geometrically thick accretion disk, without radiative compensation. In panels (a) and (b), we fix $\beta=0.01$, and the black, red, and blue curves correspond to $\sigma_{d\theta}=0.01$, $0.03$, and $0.05$, respectively, while these curves in panels (c) and (d) correspond to $\beta=0.005$, $0.015$, and $0.025$, respectively, with a fixed value of $\sigma_{d\theta}=0.025$. The second row adds the contribution of the radiative compensation on top of the first row. On the left, we fix $w_{\textrm{p}}=2$, and the curves from bottom to top correspond to $r_{\textrm{p}}=4$, 6, 8, and 10, respectively; on the right, we fix $r_{\textrm{p}}=7$, and the curves from bottom to top correspond to $w_{\textrm{p}}=0.5$, $1$, $2$, $3$, and $4$, respectively. In each panel, we fix $p_{1}=-1.5$ and $p_{2}=-0.5$. The dynamical parameters of the accretion environment are fixed to $V_{\textrm{max}}=0.2$, $p_{3}=0.5$, $\Psi=0.6$, and $\lambda=10$.}      
\label{fig18}                 
\end{figure*}

We first discuss the influence of the geometrically thick accretion disk parameters, including $\sigma_{d\theta}$, $\beta$, $r_{\textrm{p}}$, and $w_{\textrm{p}}$, on the one-dimensional intensity distribution. As shown in Fig. \ref{fig18}, we find that in one dimension, the effect of $\sigma_{d\theta}$ is relatively pronounced: increasing this parameter enhances the specific intensity of the light rays, as shown in panels (a) and (b). In contrast, the effect of $\beta$ is almost imperceptible. This is reasonable, since $\beta$ governs the broadening behavior of the disk emission in the plane containing the spin axis. Regarding the contribution of the radiative compensation, we find that changing either the position or the width of the compensation can affect the intensity distribution, manifesting as a bump in the profile. However, when the compensation region is located far from the central object or has a narrow width, its influence becomes negligible. Overall, in terms of whether the parameter produces a significant effect, $\sigma_{d\theta}$, $r_{\textrm{p}}$, and $w_{\textrm{p}}$ exhibit certain observational potential. In particular, the influence of $\sigma_{d\theta}$ remains pronounced across different inclination angles.

\begin{figure*}
\centering                   
\includegraphics[width=15cm]{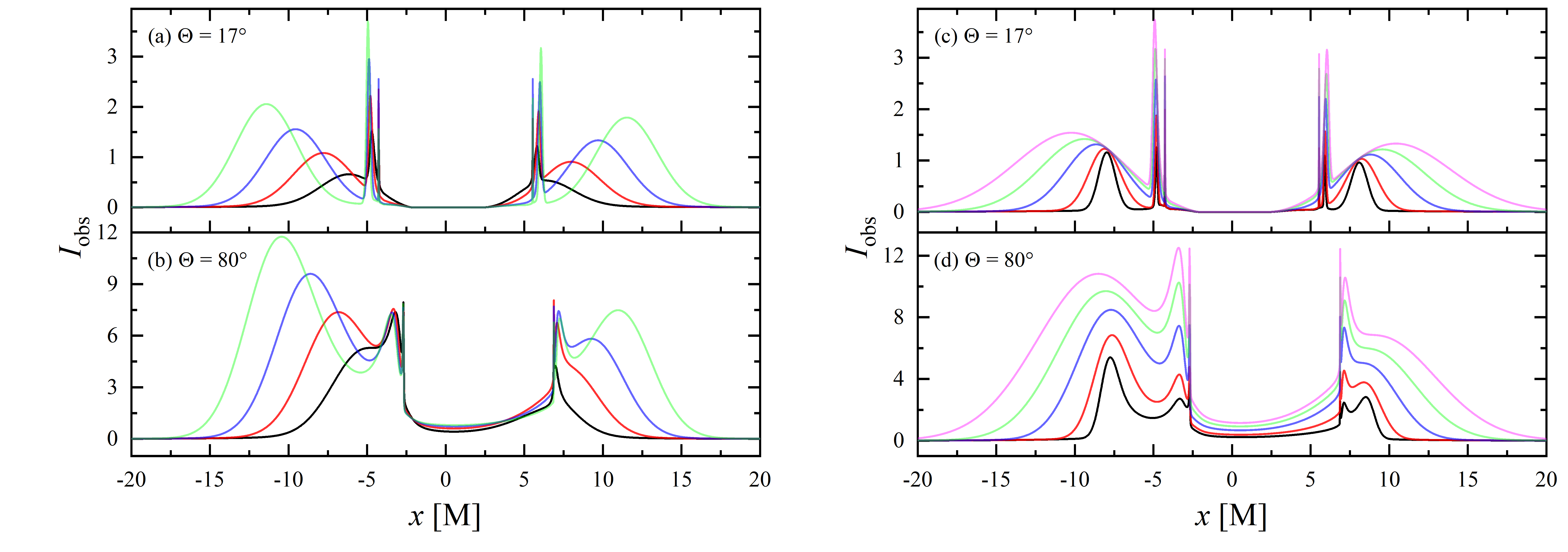}
\caption{Specific intensity distributions of light rays along the $x$-axis of the observation screen for different parameter spaces. In panels (a) and (b), we fix the Gaussian bump parameter $\sigma_{br}=2$, and the black, red, blue, and green curves correspond to $r_{\textrm{b}}=4$, $6$, $8$,  and $10$, respectively; the curves of the same colors in panels (c) and (d) correspond to $\sigma_{br}=0.5$, $1$, $2$, $3$, and $4$, respectively, with a fixed bump position $r_{\textrm{b}}=7$. Here, $\sigma_{b\theta}=\sigma_{d\theta}=0.1$, and $\beta=0.05$; the remaining model parameters are the same as in Fig. \ref{fig18}, and the radiative compensation is switched off.}      
\label{fig19}                 
\end{figure*}

Next, we focus on the influence of the Gaussian bump on the one-dimensional specific intensity, as shown in Fig. \ref{fig19}. We find that, as a global structure, the shock wave can produce an enhancement in the intensity at symmetric positions, and the intensity contrast varies with the inclination angle. Changing either the position or the width of the shock wave can significantly affect the intensity distribution. At the same time, it is worth noting that the bump parameters are unlikely to be degenerate with the observation inclination. This is because changing the observation inclination mainly affects the Doppler effect, introducing an asymmetry in the horizontal distribution of the intensity, which is fundamentally different from the effect of an axisymmetric Gaussian bump.

As global structures, the Gaussian bump and the radiative compensation are expected to be partially degenerate, since both can produce bright rings. However, the bump generally yields a higher intensity at any given position, while the compensation mainly suppresses the radial decay to sustain the emission level, and becomes increasingly ineffective at large radii.

Our model also includes localized spots. However, on a fixed horizontal line of the observation screen, assessing the influence of a source that can be placed flexibly is not exhaustive. Nevertheless, we can reasonably infer that, as a conspicuous patch in the image, the localized spot has observational potential, for instance through its position and size. Moreover, the spot parameters are unlikely to be degenerate with those of the other components of the accretion environment, because the spot is a localized feature, whereas the geometrically thick disk, the radiative compensation, and the shock wave are all global structures.
\section{Summary and conclusions}
In this paper, we have constructed, from a phenomenological perspective, a novel analytic accretion environment and an associated dynamical framework for black hole spacetimes, as detailed in Eqs. \eqref{1}--\eqref{9} and \eqref{19}--\eqref{24}. The accretion environment comprises multiple components, including a disk, an emission plateau, shock-like bumps, and localized spots, while the dynamical model adopts a global rotation approximation and allows flexible adjustment of the ratio between the radial and azimuthal velocities of the accreting material near the black hole.

Owing to the parameterized nature and the high degree of freedom of our model, it can qualitatively mimic various potential high-energy events around black holes in astrophysical contexts, such as magnetic reconnection, stream-stream collisions, and compact emission regions. The proposed analytic accretion environment offers a new perspective on modeling high-energy plasmas in curved spacetimes, and provides a convenient alternative for gaining deeper insights into accretion mechanisms.

We have applied this accretion model to Kerr black hole image simulations. We find that the simulated images also exhibit the observational characteristics of Kerr black holes, accurately reproducing the critical curve and capturing the inner shadow. However, we need to clarify that the model constructed in this paper is a parameterized model, which cannot yet be self-consistently mapped onto an astrophysical background; it can only qualitatively mimic the potential imprints of astrophysical events on the images. The advantage of a parameterized model lies in its flexible parameter space, which allows novel features to emerge in Kerr black hole images. In particular, with the inclusion of the Gaussian bump $j_{\textrm{b}}(r,\theta)$ or the localized spot $j_{\textrm{s}}(r,\theta,\varphi)$, the images exhibit multiple bright spots and bright arcs, a phenomenon that has rarely been reported in the current literature, as most existing studies consider only a single accretion disk and do not account for potential high-energy events embedded within the disk. Overall, in terms of theoretical image simulations, our accretion model can provide potential guidance for inferring the accretion environment from image features.

Interestingly, the emission around black holes may also be dominated or partly dominated by relativistic jets \cite{2021ApJ...909..168K,Papoutsis:2022kzp,Zhang:2024lsf,Shen:2023nij}. Even the EHT cannot rule out the possibility that jets serve as an emission source in black hole imaging \cite{2021ApJ...909..168K,Chavez:2024fwi,LaBella:2026say}. Therefore, it is necessary to incorporate the jet contribution into the accretion environment we have constructed. In fact, in a study nearing completion, we approximate the jet as a solid cone centered on the black hole spin axis, with its emission decaying radially and following a Gaussian distribution about the spin axis. On this basis, we have examined the influence of the jet opening angle on the potential observational features of rotating black holes. In follow-up work, we plan to introduce weight factors to further reveal the image features of multi-component emission sources under different ratios of jet to disk emission. On the other hand, in future in-depth investigations, we plan to seek astrophysical counterparts of the parameterized accretion environment constructed in this paper.
\section*{Acknowledgements}
We would like to express our sincere gratitude to the anonymous referee for the valuable comments and suggestions, which have helped us improve this manuscript. Shiyang Hu gratefully acknowledges Dr. Jiewei Huang at Peking University for constructive discussions. This work was supported by the National Natural Science Foundation of China under Grant Nos. 12403081 and 12505059, and the China Postdoctoral Science Foundation under Grant No. 2025MD784184. Guansheng He is partially funded by the Science and Technology Innovation Program of Hunan Province (Grant No. 2026RC3210), the Natural Science Foundation of Hunan Province (Grant No. 2026JJ50351), and the Scientific Research Foundation of the Hunan Provincial Education Department (Grant No. 25B0373).

\bibliographystyle{spphys}
\bibliography{references} 
\end{document}